\documentclass[prd,final,aps]{revtex4}

\usepackage{graphicx}

\begin{document}

\bibliographystyle{prsty} 

\title{The periodic standing-wave approximation: nonlinear
scalar fields, adapted coordinates, and the eigenspectral method\\
}

\author{Benjamin Bromley}
\affiliation{Department of Physics,
University of Utah, Salt Lake City, Utah 84112}  
\author{Robert Owen}
\affiliation{Theoretical Astrophysics, California
Institute of Technology, Pasadena, CA 91125}
\author{Richard H.~Price} 
\affiliation{Department of Physics \& Astronomy and Center for 
Gravitational Wave Astronomy, University of Texas at Brownsville,
Brownsville, TX 78520}

\begin{abstract}
\begin{center}
{\bf Abstract}
\end{center}

The periodic standing wave (PSW) method for the binary inspiral of
black holes and neutron stars computes exact numerical solutions for
periodic standing wave spacetimes and then extracts approximate
solutions of the physical problem, with outgoing waves. The method
requires solution of a boundary value problem with a mixed (hyperbolic
and elliptic) character.
We present here a new numerical method for such problems, based on
three innovations: (i)~a coordinate system adapted to the geometry of
the problem, (ii)~an expansion in multipole moments of these
coordinates and a filtering out of higher moments, and (iii)~the
replacement of the continuum multipole moments with their analogs for
a discrete grid.  We illustrate the efficiency and accuracy of this
method with nonlinear scalar model problems. Finally, we take
advantage of the ability of this method to handle highly nonlinear
models to demonstrate that the outgoing approximations extracted from
the standing wave solutions are highly accurate even in the presence
of strong nonlinearities.

\end{abstract}

\maketitle

\section{Introduction}\label{sec:intro} 

\subsection{Background}

The detection and interpretation of gravitational wave signals from
inspiralling black holes or neutron stars requires a solution of
Einstein's equations for the late stages of the
inspiral\cite{flanaganhughes,damouriyersath,Hughes:2001ch}.  Much effort is going
into the development of computer codes that will evolve solutions
forward in time. 
 For recent progress 
see \cite{Bruegmann:2003aw,Baumgarte:2002jm,Campanelli:2004zw}.
Such
codes will eventually provide the needed answers about the strong
field interaction and merger of the binary objects, but many technical
challenges of such a computation slow the development of the needed
codes. This has led us to propose, as a near term alternative, the
periodic standing wave (PSW) approach.  Elements of this approximation
have been introduced elsewhere \cite{WKP,WBLandP,rightapprox}, but are most
thoroughly presented in a recent paper\cite{paperI} that we will
hereafter refer to as ``Paper I.''  In the PSW approach, a numerical
solution is sought to Einstein's equations, not for a spacetime
geometry evolved from initial data, but rather for sources and fields
that rotate rigidly (i.e.\,, with a helical Killing vector) and that
are coupled to standing waves.  

Paper I gives the details of how to extract from this solution an
approximation to the problem of interest: a slowly inspiralling pair
of objects coupled to outgoing waves. Paper I also describes the
nature of the mathematical problem that must be solved numerically: a
boundary-value problem with ``standing wave boundary conditions'' on a
large sphere surrounding the sources. The differential equations of
this boundary value problem are mixed, elliptical in one region
(inside a ``light cylinder'') and hyperbolic in another (outside).

The method of solution in Paper I was straightforward. The
differential equations and boundary conditions were implemented with a
finite difference method (FDM) in a single patch of standard
corotating spherical coordinates. The equations were solved with
Newton-Raphson iteration of a sequence of linear approximations, and a
straightforward inversion of each linear approximation.  The relative
simplicity of this approach was useful to demonstrate the basic
well-posedness and solubility of the problem and to illustrate the
important issues of the PSW method, especially the ``effective
linearity'' that explains the accuracy of the PSW approximation for
the physical solution. The method, however, has severe shortcomings.
Multipole moments, and hence spherical coordinates are necessary in
the wave zone for the imposition of outer boundary conditions and for
the extraction of outgoing solutions from standing-wave solutions.
Spherical, and other standard coordinates are, however, not well
suited to resolving the relatively small sources of the binary. This
is especially true if the sources are to be represented by boundary
conditions on the outer surface of a source, rather than by explicit
source terms.  The usual technique for handling such problems is
coordinate patches and interpolation. This would be particularly
inconvenient for the PSW computations since standard iterative
approaches are inapplicable to mixed equations.

In this paper we report on an alternative approach, one that has the
disadvantage of adding some analytic complexity to the problem, and
some worrisome features. But it is a method that gives both remarkably
efficient results for model problems, and a potentially useful new
approach to the coupling of moving sources to their radiation
field. This new method is based on a coordinate system that is adapted
to the local structure of the sources and to the large-scale structure of
the distant waves.  Though the PSW computations have been 
the proximate motivation for introducing an adapted coordinate system,
the success with this system suggests that its utility may be more
broadly applicable. Such coordinates, in fact, have already been 
exploited, even in
numerical relativity.  ``\v{C}ade\v{z} coordinates''\cite{Cadez71,cadez}, a
carefully adapted coordinate system of this type, was used in much of
the work on head-on collisions of black holes, and has more recently found to 
be useful\cite{Cook91,Cook93}
for initial data and apparent horizon finding. Like the \v{C}ade\v{z}
coordinates, our coordinate systems will reduce to source-centered
spherical polar coordinates in the vicinity of the sources, and to
rotation-centered spherical polar coordinates far from the sources.

The core of the usefulness of the adapted coordinates is that the
field near the sources is well described by a few multipoles in these
coordinates, primarily the monopole of the sources, and that the field
far from the sources is well described only by a few multipoles in
these coordinates. A spectral method (that is, a multipole
decomposition), therefore, requires only a small number of
multipoles. We will demonstrate, in fact, that for mildly relativistic
sources (source velocity = 30\% $c$), excellent results are found when
we keep only monopole and quadrupole terms.

There is, of course, a price to be paid for this. For one thing, there
is additional analytic complexity in the set of equations. Another
difficulty is the unavoidable coordinate singularity that is a feature
of coordinates adapted to the two different limiting regions. Still,
the potential usefulness of the method, and the success reported 
here have led to us treating this approach as the main focus 
of our computations in the PSW work.

\subsection{Nonlinear model probelm }

The innovative features of this method present enough new
uncertainties that it is important to study this method in the context
of the simplest problem possible. We use, therefore, the same model
problem as in Paper I, a simple scalar field theory with an adjustable
nonlinearity. We will find it quite useful to set the nonlinearity to
zero for comparison with the known solution of the linear problem,
since many features of our method are unusual even for a linear
problem.  

For the description of our model problem, we start with Euclidean
space coordinatized by the usual spherical coordinates
$r,\theta,\phi$, and we consider sources concentrated near the points
$r=a$, $\theta=\pi/2$, in the equatorial plane, and moving
symmetrically according to $\phi=\Omega t$ and $\phi=\Omega t+\pi$.
As in Paper I, we seek a solution of the flat-spacetime scalar field equation
\begin{equation}\label{fieldtheory} 
\Psi_{;\alpha;\beta}g^{\alpha\beta}
+\lambda F=\nabla^2\Psi-\partial_t^2\Psi+ \lambda F={\rm Source}\,,
\end{equation}
where $F$ depends nonlinearly on $\Psi$. The explicit form of $F$
will be the same as that in Paper I. This will allow comparisons
with the results of the very different numerical technique in Paper I,
and, as in Paper I, allows a very useful comparison of the near-source
nonlinear solution with an analytic limit.

We are looking for solutions to Eq.~(\ref{fieldtheory}) with the same
helical symmetry as that of the source motions, that is, solutions for
which the Lie derivative ${\cal L}_\xi\Psi$ is zero for the Killing
vector $\xi=\partial_t+\Omega\partial_\phi$. It is useful to introduce
the auxiliary coordinate $\varphi\equiv\phi-\Omega t$. In terms of
spacetime coordinates $t,r,\theta,\varphi$ the Killing vector is simply
$\partial_t$ and the symmetry condition becomes the requirement that
the scalar field $\Psi$ is a function only of the variables $r$,
$\theta$ and $\varphi$.  (We are assuming, of course, that the form of
the nonlinear term is compatible with the helical symmetry.)  It is
useful to consider the symmetry to be equivalent to the rule
\begin{equation}\label{replacement1} 
\partial_t\rightarrow\;-\Omega\partial_\varphi
\end{equation}
for scalar functions. 
In terms of the $r$, $\theta$,
$\varphi$
variables, Eq.~(\ref{fieldtheory}) for $\Psi(r,\theta,\varphi) $ takes the
explicit form
\begin{equation}\label{origcoords} 
\frac{1}{r^2}\;\frac{\partial}{\partial r}
\left(r^2
\frac{\partial\Psi}{\partial r}
\right)
+\frac{1}{r^2\sin\theta}\frac{\partial}{\partial\theta}\left(
\sin\theta\,\frac{\partial\Psi}{\partial\theta}\right)
+\left(\frac{1}{r^2\sin^2\theta}-\Omega^2\right)
\frac{\partial^2\Psi}{\partial\varphi^2}
+\lambda F(\Psi,r,\theta,\varphi)={\rm Source}\,,
\end{equation}
that was used in Paper I. 

\subsection{Outline and summary}

The remainder of this paper has the following organization, and makes
the following points. In Sec.~\ref{sec:adapcoord} we introduce the
concept of adapted coordinates, comoving coordinates that conform to
the source geometry near the source and that become spherical comoving
coordinates far from the source. A particular system of adapted
coordinates, two center bispherical coordinates (TCBC), is introduced in
this section. Though these coordinates are not optimal for
computational accuracy, they have the advantage of analytic simplicity
and are the only adapted coordinates explicitly used in the
computations of this paper. Though the TCBC system is relatively
simple it is still sufficiently complex that that many details of the
use of this method are relegated to Appendix \ref{app:3Dcoeffs}.

In Sec.~\ref{sec:adapcoord} we also discuss the use of these adapted
coordinates in a FDM calculation, and explain the
computational difficulties we encountered in trying to find stable solutions
with this approach. These difficulties led us to use a spectral type
method with  the adapted coordintes.  In Sec.~\ref{sec:spectral} we
present the fundamental ideas of expanding the solution in spherical
harmonics of the angular adapted coordinates. In this section we
also explain why we are not, strictly speaking, using a
spectral method since we do the angular differencing by FDM, not by
relationships of the spherical harmonics. (For background on spectral 
methods, and an important recent use of spectral methods in numerical 
relativity, see \cite{Pfeiffer:2002wt}.)
Furthermore, we keep many fewer multipoles than would in principle be
justified by the number of points in our angular grid. This
``multipole filtering'' is one of the most interesting and innovative
aspects of our method. Because the adapted coordinates in some sense
handle much of the computation analytically only a few multipoles need
be kept. In most of the results, in fact, only monopole and quadrupole
moments are kept.

To illustrate a more standard spectral method, we present in Appendix
\ref{app:SSM2D} a standard spectral approach to the linear PSW problem
in two spatial dimensions described in TCBC coordinates. The appendix
also uses severe multipole filtering and serves to demonstrate in a
very different, and generally simpler, numerical context the
fundamental correctness of multipole filtering.

For the problem in three spatial dimensions, we have found that a
 special technique must be used for multipole expansion and
 multipole filtering.  A straightforward approach would use the
 continuum multipoles evaluated on the angular grid. We explain in
 Sec.~\ref{sec:spectral} why this method involves unacceptably large
 numerical errors, and why we introduce a second innovative numerical
 technique, one that we call the ``eigenspectral method.'' In place of
 the continuum spherical harmonics evaluated on the angular grid, we
 use eigenvectors of the angular FDM Laplacian. These eigenvectors
 approach the grid-evaluated continuum spherical harmonics as the grid
 becomes finer but, as we explain in this section, the small
 differences are very important in the multipole expansion/filtering
 method. Some of the details of the eigenspectral method are put into
 Appendix \ref{app:ESMdetails}, in particular the way the FDM angular
 Laplacian can be treated as a self-adjoint operator.

Section \ref{sec:modmeth} starts by presenting the details of the
model scalar field problems to which we apply the eigenspectral
method: the choice of the nonlinearity, and the justification for this
choice; the manner in which we choose data on an inner boundary taken
to be the outer surface of a source; the outer radiative boundary
conditions; the Newton-Raphson procedure for finding solutions to
nonlinear problems; and the method by which we extract approximate
nonlinear outgoing solutions from computed nonlinear standing wave
solutions.

This is followed, in Sec.~\ref{sec:numresults}, by a presentation of
numerical results that demonstrate convergence of the method. These
results show that the numerical methods are quite accurate despite the
inclusion of only a very minimal number of multipoles. In addition,
the power of the numerical method allows us to compute models with
much stronger nonlinearity than could be handled with the
straightforward FDM of Paper I. For these highly nonlinear models we
confirm the ``effective linearity'' that was demonstrated in Paper I
with less dramatic models: the outgoing solution extracted from a
standing wave solution is an excellent approximation to the true
outgoing solution, even for very strong nonlinearity.
Conclusions are briefly summarized in Sec.~\ref{sec:conc}.

Throughout this
paper we follow the notation of Paper I\cite{paperI}. (A few changes
from the notation and choices of Paper I are made to correct minor
errors of Paper~I: {(i)}~the point source delta function is now
divided by a Lorentz $\gamma $ factor, as explained following
Eq.~(\ref{oursigma}). {(ii)}~The nonlinearity parameter $\lambda$
was used with inconsistent dimensionality in Paper~I. Here $\lambda$
is consistently treated as a dimensionless parameter, requiring the
insertion of a factor $1/a^2$ in the model nonlinearity of
Eq.~(\ref{modelF}).

\section{Adapted coordinates}\label{sec:adapcoord} 

\subsection{General adapted coordinates}
For the definition of the adapted coordinates it is useful to introduce
several Cartesian coordinate systems. We shall use the notation $x,y,z$
to denote inertial Cartesian systems related to $r,\theta,\phi$ in the 
usual way (e.g.\,, $z$ is the rotation axis, one of the source points
moves as $x=a\cos{(\Omega t)}$, $y=a\sin{(\Omega t)}$, 
and so forth). We now introduce a 
comoving Cartesian system $\widetilde{x},\widetilde{y},\widetilde{z}$
by
\begin{equation}\label{sphericals} 
\widetilde{z}=r\cos\theta\quad\quad
\widetilde{x}=r\sin\theta\cos{(\phi-\Omega t)}\quad\quad
\widetilde{y}=r\sin\theta\sin{(\phi-\Omega t)}\,.
\end{equation}
\begin{figure}[ht] 
\includegraphics[width=.25\textwidth]{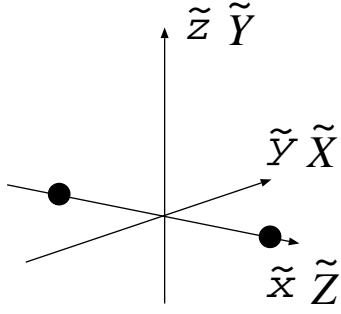}
\caption{Two sets of comoving Cartesian coordinates.
\label{fig:XYZ}}
\end{figure}
In this system, as in the inertial  $x,y,z$ system, the $\widetilde{z}$ axis
is the azimuthal axis. We next define the comoving 
system 
\begin{equation}\label{cartrelns} 
\widetilde{X}=\widetilde{y}\quad\quad
\widetilde{Y}=\widetilde{z}\quad\quad
\widetilde{Z}=\widetilde{x}\,,
\end{equation}
in which the azimuthal $\widetilde{Z}$ axis is not the rotation axis,
but rather is the line through the source points, as shown in
Fig.~\ref{fig:XYZ}.

Our goal now is to introduce new comoving coordinates
$\chi(\widetilde{X},\widetilde{Y},\widetilde{Z})$,
$\Theta(\widetilde{X},\widetilde{Y},\widetilde{Z})$,
$\Phi(\widetilde{X},\widetilde{Y},\widetilde{Z})$ that are better
suited to a description of the physical problem, and that allow for
more efficient computation.  We will assume that the coordinate
transformation is invertible, except at a finite number of discrete
points, so that we may write
$\widetilde{X},\widetilde{Y},\widetilde{Z}$, or
$\widetilde{x},\widetilde{y},\widetilde{z}$ as functions of
$\chi,\Theta,\Phi$.

In terms of the comoving Cartesian coordinates, the helical symmetry
rule in Eq.~(\ref{replacement1}) takes the form
\begin{equation}\label{replacement2} 
\partial_t\rightarrow-\Omega\left(
\widetilde{x}\frac{\partial}{\partial \widetilde{y}}
-\widetilde{y}\frac{\partial}{\partial \widetilde{x}} 
\right)
=
-\Omega\left(
\widetilde{Z}\frac{\partial}{\partial\widetilde{X}}
-\widetilde{X}\frac{\partial}{\partial\widetilde{Z}}\ 
\right)\ .
\end{equation}
Our nonlinear scalar field equation  of Eq.~(\ref{fieldtheory})
can then be written, for helical symmetry, as
\begin{equation}\label{3dfieldtheory} 
{\cal L}\Psi+\lambda F=\frac{\partial^2\Psi}{\partial \widetilde{X}^2}+
\frac{\partial^2\Psi}{\partial\widetilde{Y}^2}+
\frac{\partial^2\Psi}{\partial\widetilde{Z}^2}
-\Omega^2
\left(\widetilde{Z}\frac{\partial}{\partial\widetilde{X}}
-\widetilde{X}\frac{\partial}{\partial\widetilde{Z}}\right)^2\Psi
+ \lambda F={\rm Source}\ .
\end{equation}
This field equation can be expressed completely
in terms of adapted coordinates in the form
\begin{displaymath}
{\cal L}\Psi+\lambda F=A_{\chi\chi}\;\frac{\partial^2\Psi}{\partial\chi^2}
+A_{\Theta\Theta}\;\frac{\partial^2\Psi}{\partial\Theta^2}
+A_{\Phi\Phi}\;\frac{\partial^2\Psi}{\partial\Phi^2}
+2A_{\chi\Theta}\;\frac{\partial^2\Psi}{\partial\chi\partial\Theta}
+2A_{\chi\Phi}\;\frac{\partial^2\Psi}{\partial\chi\partial\Phi}
+2A_{\Theta\Phi}\;\frac{\partial^2\Psi}{\partial\Theta\partial\Phi}
\end{displaymath}
\begin{equation}\label{waveq} 
+B_{\chi}\;\frac{\partial\Psi}{\partial\chi}
+B_{\Theta}\;\frac{\partial\Psi}{\partial\Theta}
+B_{\Phi}\;\frac{\partial\Psi}{\partial\Phi}+\lambda F={\rm Sources}\ .
\end{equation}

It is straightforward to show that the
$A_{ij}$ and $B_{i}$ coefficients here are 
given by
\begin{eqnarray}
A_{\chi\chi}&=&\vec{\nabla}\chi\cdot\vec{\nabla}\chi
-{\Omega^2}\bar{A}_{\chi\chi}\label{Achichi}  \\
A_{\Theta\Theta}&=&\vec{\nabla}\Theta\cdot\vec{\nabla}\Theta
-{\Omega^2}\bar{A}_{\Theta\Theta}\\
A_{\Phi\Phi}&=&\vec{\nabla}\Phi\cdot\vec{\nabla}\Phi
-{\Omega^2}\bar{A}_{\Phi\Phi}\\
A_{\chi\Theta}&=&\vec{\nabla}\chi\cdot
\vec{\nabla}\Theta-{\Omega^2}\bar{A}_{\chi\Theta}\\
A_{\chi\Phi}&=&\vec{\nabla}\chi\cdot
\vec{\nabla}\Phi-{\Omega^2}
\bar{A}_{\chi\Phi}\\
A_{\Theta\Phi}&=&\vec{\nabla}\Theta\cdot
\vec{\nabla}\Phi-{\Omega^2}
\bar{A}_{\Theta\Phi}\\
B_{\chi}&=&{\nabla^2}\chi
-{\Omega^2}\bar{B}_{\chi}\\
B_{\Theta}&=&{\nabla^2}\Theta
-{\Omega^2}\bar{B}_{\Theta}\\
B_{\Phi}&=&{\nabla^2}\Phi
-{\Omega^2}\bar{B}_{\Phi}\label{BPhi}  \ .
\end{eqnarray}
Here the gradients, Laplacians and dot products are to be taken 
treating the 
$\widetilde{X}$,$\widetilde{Y}$,$\widetilde{Z}$
as Cartesian coordinates, so that, for example,
\begin{equation}
\vec{\nabla}\chi\cdot\vec{\nabla}\Theta=
\frac{\partial\chi}{\partial\widetilde{X}}
\frac{\partial\Theta}{\partial\widetilde{X}}
+\frac{\partial\chi}{\partial\widetilde{Y}}
\frac{\partial\Theta}{\partial\widetilde{Y}}
+\frac{\partial\chi}{\partial\widetilde{Z}}
\frac{\partial\Theta}{\partial\widetilde{Z}}\,.
\end{equation}
The form of the $\bar{A}_{ij} $ and $\bar{B}_j $ terms in
Eqs.~(\ref{Achichi})--(\ref{BPhi}) 
are given, for general adapted coordinates, in 
Eqs.~(\ref{barAchichi}) --
(\ref{deq}).

\subsection{A specific adapted coordinate system: TCBC}

\begin{figure}[ht] 
\includegraphics[width=.35\textwidth]{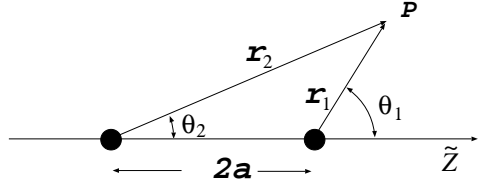}
\caption{Geometric basis for the TCBC adapted coordinates.\label{fig:2Ddef}}
\end{figure}

Before discussing general features of an adapted coordinate system, it
will be useful to give a specific example.  For that example, we
choose a coordinate system $\chi,\Theta,\Phi$ that is particularly
simple in form, though (as we shall discuss below) not the choice that
is numerically most efficient.  The chosen coordinates are most easily
understood by starting with the distances $r_1$ and $r_2$ from the
source points, and with the angles $\theta_1$, $\theta_2$ shown in
Fig.~\ref{fig:2Ddef}.  The formal definitions of the adapted
coordinates are
\begin{eqnarray}
\chi&\equiv&\sqrt{r_1r_2}
=\left\{\left[\left(\widetilde{Z}-a\right)^2+\widetilde{X}^2
+\widetilde{Y}^2\right]
\left[\left(\widetilde{Z}+a\right)^2+\widetilde{X}^2
+\widetilde{Y}^2\right]\right\}^{1/4}\label{chiofXYZ}\\
\Theta&\equiv&\frac{1}{2}\left(\theta_1+\theta_2\right)=
\frac{1}{2}\tan^{-1}\left(\frac{2\widetilde{Z}
\sqrt{\widetilde{X}^2+
\widetilde{Y}^2
\;}}{
\widetilde{Z}^2-a^2-\widetilde{X}^2-\widetilde{Y}^2
}\right)\label{ThetofXYZ}\\
\Phi&\equiv&\tan^{-1}{\left(\widetilde{X}/\widetilde{Y}\right)}
\label{PhiofXYZ}\ .
\end{eqnarray}
This choice is sometimes called ``two-center bipolar
coordinates''\cite{bipolar}, hereafter TCBC, and is equivalent to the
zero-order coordinates used by \v{C}ade\v{z}\cite{Cadez71,cadez}.
\begin{figure}[ht]

\hspace{-2.6in}\includegraphics[width=.41\textwidth]{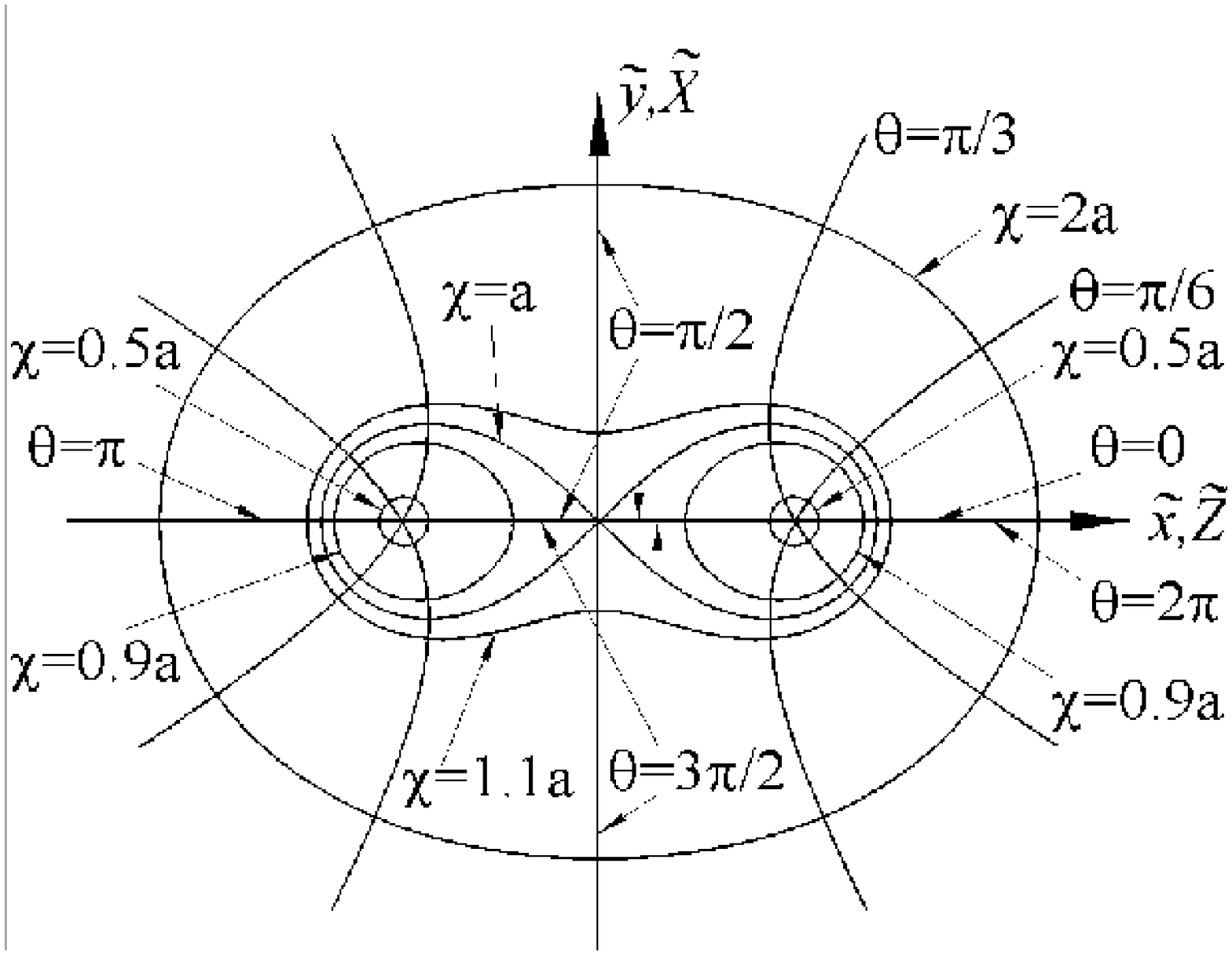}

\vspace{-2.35in}

\hspace{2.9in}\includegraphics[width=.33\textwidth]{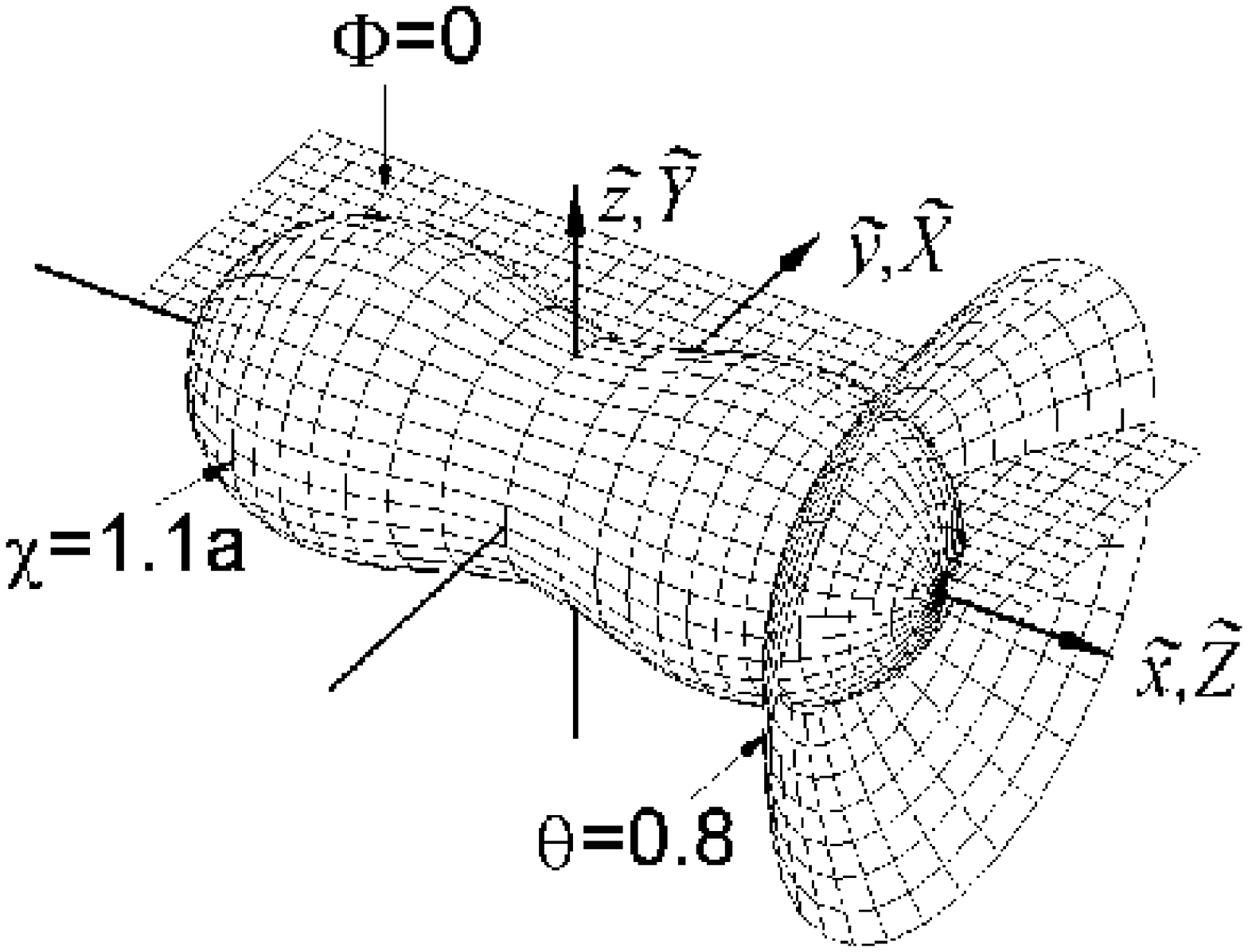}

\caption{
Adapted coordinates in the 
$\widetilde{x},\widetilde{y}$ plane, and  three-dimensional coordinate
surfaces.
\label{fig:2n3Ros}}
\end{figure}

An attractive feature of this particular
choice of adapted coordinates is that the above relationships can be
inverted in simple closed form to give
\begin{eqnarray}
\widetilde{Z}&=&\sqrt{\frac{1}{2}\left[
a^2+\chi^2\cos{2\Theta}+\sqrt{\left(a^4+2a^2\chi^2\cos{2\Theta}+\chi^4\right)\  }
\right]\ }\label{zof}\\
\widetilde{X}&=&\sqrt{\frac{1}{2}
\left[
-a^2-\chi^2\cos{2\Theta}+
\sqrt{\left(a^4+2a^2\chi^2\cos{2\Theta}+\chi^4\right)\  }
\right]\ }\;\cos{\Phi}\ .\label{xof}\\
\widetilde{Y}&=&\sqrt{\frac{1}{2}
\left[
-a^2-\chi^2\cos{2\Theta}+
\sqrt{\left(a^4+2a^2\chi^2\cos{2\Theta}+\chi^4\right)\  }
\right]\ }\;\sin{\Phi}\ .\label{yof}
\end{eqnarray}
The meaning of the $\chi,\Theta$ coordinates in the
$\widetilde{x},\widetilde{y}$ plane (the 
$\widetilde{Z},\widetilde{X}$ plane) is shown on the
left in Fig.~\ref{fig:2n3Ros}; a picture of three-dimensional $\chi$, $\Theta$,
and $\Phi$ surfaces is shown on the right.

The geometrical definition inherent in Fig.~\ref{fig:2Ddef}
suggests that the adapted coordinate surfaces have 
the correct limit far from the sources. This is confirmed
by the limiting forms  Eqs.~(\ref{zof})--(\ref{yof})
for $\chi\gg a$. Aside from fractional corrections 
of order $a^2/\chi^2$ the relations  are
\begin{equation}
\widetilde{Z}\rightarrow\chi\cos{\Theta}\quad\quad
\widetilde{X}\rightarrow\chi\sin{\Theta}\cos{\Phi}\quad\quad
\widetilde{Y}\rightarrow\chi\sin{\Theta}\sin{\Phi}\ .
\end{equation}
Near the source point at $\widetilde{Z}=\pm a$, the limiting forms, aside
from fractional corrections of order $\chi^2/a^2$, are
\begin{equation}
\widetilde{Z}\rightarrow \pm a+\frac{\chi^2 }{2a}\cos{(2\Theta)}\quad\quad
\widetilde{X}\rightarrow \frac{\chi^2 }{2a}\sin{(2\Theta)}\cos\Phi\quad\quad
\widetilde{Y}\rightarrow \frac{\chi^2 }{2a}\sin{(2\Theta)}\sin\Phi\,.
\end{equation}
These limits, and Fig.~\ref{fig:2Ddef}, show that near the source point
at $\widetilde{Z}=a$ the expression $\chi^2/2a$ plays the role of
radial distance, and $2\Theta$ plays the role of polar
coordinate. (Near the source point at $\widetilde{Z}=-a$, the
expression $\chi^2/2a$ again plays the role of radius, but
the polar angle is $\pi-2\Theta$.) Notice that both for the near and the
far limit, the polar angle is defined with respect to the line through
the sources, the $\widetilde{Z}$ axis, not with respect to the
rotational $\widetilde{z}$ axis.

It is clear that our new system has a coordinate singularity at the
origin. Indeed, there must be a coordinate singularity in any such
adapted coordinate system. The switch from the small-$\chi$ coordinate
surfaces, disjoint 2-spheres around the sources, to the large-$\chi$
single 2-sphere cannot avoid a singularity. 

The remaining specification needed is the outer boundary conditions on some
large approximately spherical surface $\chi=\chi_{\rm max}$.  
For the monopole moment of the field this condition is simply that
the field dies off as $1/\chi$. For the radiative part of the field
we use the
usual Sommerfeld outgoing outer boundary
condition $\partial_t\psi=-\partial_r\psi$, approximated as
$\partial_t\psi=-\partial_\chi\psi$. The fractional error introduced
by this substitution is of order $a^2/\chi^2$. The Sommerfeld
condition itself is accurate only up to order $({\rm
wavelength}/r)$. Since the wavelength is larger than $a$, our
substitution $r\rightarrow\chi$ in the outer boundary condition
introduces negligibly small errors.  To apply the helical symmetry we
use the replacement rule in Eq.~(\ref{replacement2}) and the outgoing
boundary condition becomes
\begin{equation}\label{FDMoutbc} 
\frac{\partial\Psi}{\partial\chi}=\Omega\left(
\widetilde{Z}\frac{\partial\Psi}{\partial\widetilde{X}}
-\widetilde{X}\frac{\partial\Psi}{\partial\widetilde{Z}}\ 
\right)=\Omega\left(\Gamma^\Theta\frac{\partial\Psi}{\partial\Theta}
+\Gamma^\Phi\frac{\partial\Psi}{\partial\Phi}
+\Gamma^\chi\frac{\partial\Psi}{\partial\chi}
\right)\,,
\end{equation}
where the $\Gamma$s are given explicitly in Appendix~\ref{app:3Dcoeffs}.
 At large $\chi$ the outgoing condition can 
be written
\begin{equation}\label{FDMoutbcapprox} 
\frac{\partial\Psi}{\partial\chi}
=\Omega\,\left(\cos\Phi\frac{\partial\Psi}{\partial\Theta}
-\frac{\cos\Theta}{\sin\Theta}\,\sin\Phi
\frac{\partial\Psi}{\partial\Phi}
\right) \left(
1+{\cal O}(a^2/\chi^2)
\right)\,.
\end{equation}
The correction on the right is higher-order at the outer boundary
$\chi=\chi_{\rm max}$ and can be ignored.
The ingoing boundary condition follows by changing the sign of the right
hand side of Eq.~(\ref{FDMoutbc}) or (\ref{FDMoutbcapprox}).

The problem of Eqs.~(\ref{waveq}) and
(\ref{FDMoutbcapprox}) is a well-posed boundary-value problem
analagous to that in Paper I\cite{paperI}. As in Paper I, this problem
can be numerically implemented using the finite difference method
(FDM) of discretizing derivatives. The difference between such a
computation and that of Paper I is, in principle, only in the
coordinate dependence of the coefficients ($A_{\chi\chi}$,
$A_{\Theta\Theta},\cdots$) that appear  in the differential equation and 
($\Gamma^\Theta,\cdots
$) in the 
outer boundary condition.

\subsection{Requirements for adapted coordinates}
For the scalar problem, there are obvious advantages of the coordinate
system pictured in Fig.~\ref{fig:2n3Ros}. First, the surfaces of
constant $\chi$ approximate the surfaces of constant $\Psi$ near the
sources, where field gradients are largest, and where numerical
difficulties are therefore expected.  Since the variation with respect
to $\Theta$ and $\Phi$ is small on these surfaces, finite differencing
of $\Theta$ and $\Phi$ derivatives should have small truncation
error. The steep gradients in $\chi$, furthermore, can be dealt with
in principle by a reparameterization of $\chi$ to pack more grid
zones near the source points. An additional, independent advantage to
the way the coordinates are adapted to the source region is that these
coordinates are well suited for the specification of inner boundary
conditions on a constant $\chi$ surface. Because of these advantages
we shall reserve the term ``adapted'' to a coordinate system for which
constant $\chi$ surfaces near the source approximate spheres
concentric with the source.

A second feature of the TCBC coordinates that we shall also require in
general, is that in the region far from the sources,
$\chi$,$\Theta$,$\Phi$ asymptotically approach spherical coordinates,
the coordinates best suited for describing the radiation field.
If the approach to spherical coordinates
is second-order in $a/r$, then the outgoing boundary conditions will
be that in Eq.~(\ref{FDMoutbcapprox}).

There are practical considerations that also apply to the choice of
adapted coordinates.  The coefficients of the rotational terms in the
equation (i.e.\,, those involving $\bar{A}_{ij}$ and $\bar{B}_{i}$ in 
Eqs.~(\ref{Achichi})--(\ref{BPhi}))
require computing second derivatives of the transformation from
Cartesian to adapted coordinates. If those relationships are only
known numerically, these second derivatives will tend to be noisy.
For that reason, a desirable and perhaps necessary feature of the
adapted coordinates is  that closed form expressions exist
for $\chi( \widetilde{x},\widetilde{y},\widetilde{z})$, and $\Theta(
\widetilde{x},\widetilde{y},\widetilde{z})$. (The expression for
$\Phi$, the azimuthal angle about the line through the source points,
is trivial.) It is possible in principle, of course, to have the
adapted coordinates defined without respect to the Cartesians. In the
scalar model problem, the coordinates could be defined by giving the
form of the flat spacetime metric in these coordinates. The nature of
the helical Killing symmetry, analogous to Eq.~(\ref{replacement2})
would still have to be specified of course. The choice of 
adapted coordinates becomes a much richer subject in the 
case of the gauge-fixed general relativity problem that
is the ultimate goal of the work; see Ref.~\cite{coordpap}.

The TCBC coordinates satisfy all the practical requirements of an
adapted coordinate system. In particular, the functions $\chi(
\widetilde{x},\widetilde{y},\widetilde{z})$, and $\Theta(
\widetilde{x},\widetilde{y},\widetilde{z})$, as well as their inverses,
are all explicitly known in terms of elementary functions.  Though the
TCBC coordinates are therefore convenient, in addition to being well
suited to the problem in Eq.~(\ref{fieldtheory}), they are not
optimal. The perfect coordinates would  be those for which the
constant $\chi$ surfaces agree exactly with the constant $\Psi$
surfaces. This of course is impossible in practice (and, in addition,
would not be compatible with the requirement that the coordinates go
asymptotically to spherical coordinates). 
We should therefore modify the criterion for the ``perfect
coordinates'' to that of having $\Psi$ constant on constant
$\chi$ surfaces for no rotation ($\Omega=0$).  The TCBC coordinate
system, in fact, does satisfy that requirement for the version of the
problem of Eq.~(\ref{fieldtheory}) in two spatial dimensions with no
nonlinearity, as detailed in Appendix~\ref{app:SSM2D}.  Due to this
``near perfection'' of the TCBC coordinates for the linear
two-dimensional problem we found that we were able to achieve very
good accuracy for that case with moderate rates of rotation.

These considerations suggest that we could achieve an improvement
over the TCBC coordinates, by choosing $\chi$ to be proportional to
solutions of the nonrotating case of Eq.~(\ref{fieldtheory}) in three
spatial dimensions. Since the nonlinear case would result in a
solution that is known only numerically, we can follow the pattern of
the two-dimensional case and choose $\chi$ simply to be proportional
to the solution of the linear nonrotating three-dimensional problem.
The $\Theta$ coordinate that is orthogonal to this $\chi$ would have
to be found numerically, and would therefore be troublesome. But there
is no need for $\Theta$ and $\chi$ to be orthogonal. We could,
therefore, use the TCBC definition of $\Theta$ in
Eq.~(\ref{ThetofXYZ}). An improved set of adapted coordinates, then,
would seem to be
\begin{equation}\label{proposed_adaps} 
\chi=\frac{1}{2}\,\left(
\frac{1}{r_1}+\frac{1}{r_2}
\right)
\quad\quad
\Theta=\frac{1}{2}\left(\theta_1+\theta_2\right)\,,
\end{equation}
where $r_i,\theta_i$ are the distances and angles shown
Fig.~\ref{fig:2Ddef}.

In this paper, we shall report only numerical results from the
simplest adapted coordinate system to implement, the TCBC
coordinates. There are two reasons for this. The first is the obvious
advantages of working with the simplicity of the TCBC case, and the
advantage of having simple explicit expressions for all coefficients
in Eq.~(\ref{waveq}).  The second reason that we do not use the
apparently superior adapted coordinate in Eq.~(\ref{proposed_adaps}),
is that we do not expect there to be an equivalent for the general
relativity problem. In that case there will be several different
unknown fields to solve for, and there is no reason to think that the
optimal coordinate system for one of the fields will be the same for
the others.

\section{Spectral methods with adapted coordinates}\label{sec:spectral} 

\begin{figure}[ht] 
\includegraphics[width=.35\textwidth]{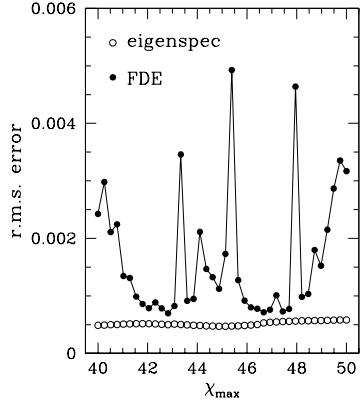}
\caption{
Error in the computed outgoing linear solution as a function of the
location of the outer boundary.  Results are shown both for
straightforward FDM in adapted coordinates and for the eigenspectral
method, explained in the text, with only monopole and quadrupole terms
kept.
\label{fig:rezcav}}
\end{figure}

The wave equation in Eq.~(\ref{waveq}), along with the boundary
conditions Eq.~(\ref{FDMoutbcapprox}), can in principle be solved by
imposing a $\chi,\Theta,\Phi$ grid and by using FDM. In practice,
numerical problems hinder a straightforward finite difference
computation. Evidence for this is shown in Fig.~\ref{fig:rezcav}, in
which the error (the difference between the computed solution and the
analytic solution) for the linear outgoing problem is plotted for
different locations of the outer boundary $\chi_{\rm max}$.  As
Fig.~\ref{fig:rezcav} shows, the quality of our solutions was highly
sensitive to small changes in grid parameters, such as the location of
the outer boundary. We attribute these difficulties to the orientation
of the finite difference grid at large distance from the
source. Loosely speaking, the spherical polar grid is ``aligned'' with
the solutions, and errors are distributed evenly on the grid.  Adapted
coordinates become spherical polar at large distances, but the polar
axis is aligned with the sources, not with the rotation axis. The
result may be a nonuniform distribution of errors, which effectively
excites spurious modes analogous to modes excited inside a resonant
cavity.

An attractive alternative  to FDM
 is to expand $\Psi$ in a complete set of functions of the angular
 coordinates $\Theta$ and $\Phi$.  Since one of our goals is to
 describe the radiation in the weak wave zone, and since $\Theta$ and
 $\Phi$ approach comoving spherical coordinates in the weak wave zone,
 the natural set of basis functions is the spherical harmonics
 $Y_{\ell m}(\Theta,\Phi)$. In terms of these we would look for a
 solution of Eq.~(\ref{waveq}) in the form
\begin{equation}\label{expansion} 
\Psi=
\sum_{\mbox{even $\ell$}}
\;
\sum_{m}a_{\ell m}(\chi)
Y_{\ell m}(\Theta,\Phi)\,.
\end{equation}
(The odd $\ell$s are omitted due to the symmetry of the problem.)

The possibility of such a spectral method has been introduced in Paper
I as a potentially powerful way of dealing with radiation from moving
sources. The reason for this is that near the source points the field
is nearly spherically symmetric, and hence can be described with very
few multipoles. Far from the source, 
the contribution from multipoles
of order $\ell$ scale  as $(a\Omega)^\ell$, so 
the radiation field is dominated
by the monopole and quadrupole, 
and again can be described with very
few multipoles. It is, therefore, plausible that with very few 
multipoles --- perhaps only the monopole and quadrupole --- the 
fields everywere can be described with reasonable accuracy.

In the multipole method, the expansion in Eq.~(\ref{expansion})
is substituted in Eq.~(\ref{waveq}) to give 
\begin{displaymath}
{\cal L}\Psi= \sum_{\ell m}\ \frac{d^2 a_{\ell m}}{d\chi^2}\;\left[A_{\chi\chi}
Y_{\ell m}\right]
+a_{\ell m}(\chi)\;
\left[A_{\Theta\Theta} \frac{\partial^2Y_{\ell m} }{\partial\Theta^2}   
+A_{\Phi\Phi} \frac{\partial^2Y_{\ell m} }{\partial\Phi^2}   
+2A_{\Theta\Phi} \frac{\partial^2Y_{\ell m} }{\partial\Theta\partial\Phi}   
+B_{\Theta} \frac{\partial Y_{\ell m} }{\partial\Theta}   
+B_{\Phi} \frac{\partial Y_{\ell m} }{\partial\Phi}   
\right]
\end{displaymath}
\begin{equation}\label{SphHarm1} 
+\frac{da_{\ell m}}{d\chi}\left[
2A_{\chi\Theta} \frac{\partial Y_{\ell m} }{\partial\Theta} 
+2A_{\chi\Phi} \frac{\partial Y_{\ell m} }{\partial\Phi} 
+B_{\chi} Y_{\ell m}
\right]\ .
\end{equation}

The next step is to project out ordinary differential equations.  This
is most naturally done by multiplying by some weight function
$W(\chi,\Theta)$ and by $Y_{\ell' m'}$, and by integrating over all
$\Theta$ and $\Phi$. The result is our multipole equations
\begin{equation}\label{radialODE} 
\sum_{\ell m}\alpha_{\ell'm'\ell m}\frac{d^2a_{\ell
m}(\chi)}{d\chi^2} +\beta_{\ell'm'\ell m}a_{\ell m}(\chi)
+\gamma_{\ell'm'\ell m}\frac{d a_{\ell
m}(\chi)}{d\chi}=S_{\ell m}\ ,
\end{equation}
where $S_{\ell m}$ is the multipole of the source term, and
where
\begin{eqnarray}
\alpha_{\ell'm'\ell m}&=&\int_0^{2\pi}d\Phi\int_0^{\pi}d\Theta\
 W(\chi,\Theta)\;
Y_{\ell' m'}^*(\Theta,\Phi)  {A}_{\chi\chi} Y_{\ell m}(\Theta,\Phi)\nonumber\\
\beta_{\ell'm'\ell m}&=&\int_0^{2\pi}d\Phi\int_0^{\pi}d\Theta\ 
W(\chi,\Theta)\;
Y_{\ell' m'}^*(\Theta,\Phi)\left[
{A}_{\Theta\Theta}\frac{\partial^2Y_{\ell m}}{\partial\Theta^2} 
+
{A}_{\Phi\Phi}\frac{\partial^2Y_{\ell m}}{\partial\Phi^2} 
+
2{A}_{\Theta\Phi}\frac{\partial^2Y_{\ell m}}{\partial\Phi\partial\Theta} 
+
{B}_{\Theta}\frac{\partial Y_{\ell m}}{\partial\Theta} 
+
{B}_{\Phi}\frac{\partial Y_{\ell m}}{\partial\Phi}
\right]\nonumber\\ 
\gamma_{\ell'm'\ell m}&=&\int_0^{2\pi}d\Phi\int_0^{\pi}d\Theta\ 
W(\chi,\Theta)\;
Y_{\ell' m'}^*(\Theta,\Phi)\left[
2{A}_{\chi\Theta}\frac{\partial Y_{\ell m}}{\partial\Theta} 
+2{A}_{\chi\Phi}\frac{\partial Y_{\ell m}}{\partial\Phi} 
+{B}_{\chi}Y_{\ell m}
\right]\label{alphbetgam}\ .
\end{eqnarray}

The problem with this straightforward approach to multipole
decomposition is that the angular integrals needed for the projection
are very computationally intensive, and the  solutions of the
differential equations in $\chi$ are very sensitive to the values of
the $\alpha$'s, $\beta$'s, and $\gamma$'s, that are computed by these
projection integrals. These shortcomings do not apply to the
2-dimensional version of the helically symmetric wave equation.  In
that case the projection integrals involve only a single integration
variable, and it proves to be fairly easy to compute accurate angular
integrals. We present the straightforward 2-dimensional multipole
expansion in Appendix \ref{app:SSM2D}. This is meant to illustrate the
multipole expansion in a particularly simple context, but more
important it demonstrates a crucial point, that we can get excellent
accuracy by keeping only two multipoles.  This 2-dimensional
computation also illustrates the alternative definiton of standing
waves, that of minimum wave amplitude, as sketched in
Paper I.

It turns out that for the 3-dimensional problem, even with only a
small number of multipoles, there are two classes of severe
computational difficulties. First, the projection integrals in
Eq.~(\ref{alphbetgam}) are very computationally intensive, especially
due to the singularity at $\Theta=\pi/2$ for $\chi/a\leq1$, a
singularity that must be canceled in the projection integrals by the
choice of the weight function $W(\chi,\Theta) $. In trials with the
linear problem, and in comparisons with the known exact answer, we
have found that accuracy of the computed field is poor unless the
integrals are done very precisely.  A second, quite distinct,
difficulty is related to the projection at the outer boundary. An
outgoing boundary condition is applied to $a_{\ell m}$, for
$\ell>0$. The radiative moments, however, are much smaller than the
monopole moment $a_{00}$.  Projection of a $a_{\ell m}$ moment with
$\ell>0 $ will be contaminated by the much larger monopole moment
$a_{00}$, due to small numerical inaccuracies in the projection.
We have found this to be a problem even in the simplest
(static linear) models.

We have used an alternative approach to multipole decomposition and
multipole filtering, an approach that gives excellent results for the
nonlinear scalar models and promises to be similarly useful in
gravitational models.   Underlying this approach is the concept that
the angular nature of the multipole components of the radiation field
is determined by FDM operations, in particular by the FDM
implementation of the Laplacian. The properties of the spherical
harmonics that make them useful in the continuum description of
radiation is taken over, in FDM computations, by the eigenvectors of
the FDM Laplacian. 

To implement this idea  we start by viewing the grid values of 
the scalar field $\Psi$ on a constant-$\chi$ surface as a vector ${\bf\Psi}$
whose components are most conveniently expressed with a double index
\begin{equation}
\Psi_{ij}=\Psi(\Theta_i,\Phi_j)\,.
\end{equation}
Here $\Theta_i$ and $\Phi_j$ are the values on the  $\Theta,\Phi$
grid with spacings $\Delta\Theta$ and $\Delta\Phi$. It follows that
${\bf\Psi} $ is a vector in a space of dimension $N\equiv
n_\Theta\times n_\Phi$.

In the $\Theta,\Phi$ continuum, the angular part of the Laplacian
at $\chi/a\gg1$ is the operator 
\begin{equation}\label{angLap} 
\nabla_{\rm ang}^2
\equiv
\frac{1}{\sin\Theta}\,\frac{\partial}{\partial\Theta}
\left[\sin\Theta\,\frac{\partial}{\partial\Theta}
\right]+\frac{1}{(\sin\Theta)^2
}\,\frac{\partial^2
}{\partial\Phi^2}\,.
\end{equation}
In a FDM this  is replaced by an operator in the $N$-dimensional space
of angular grid values. Our eigenspectral method is based on finding
the eigenvectors of this $N$-dimensional operator.

\begin{figure}[ht] 
\includegraphics[width=.45\textwidth]{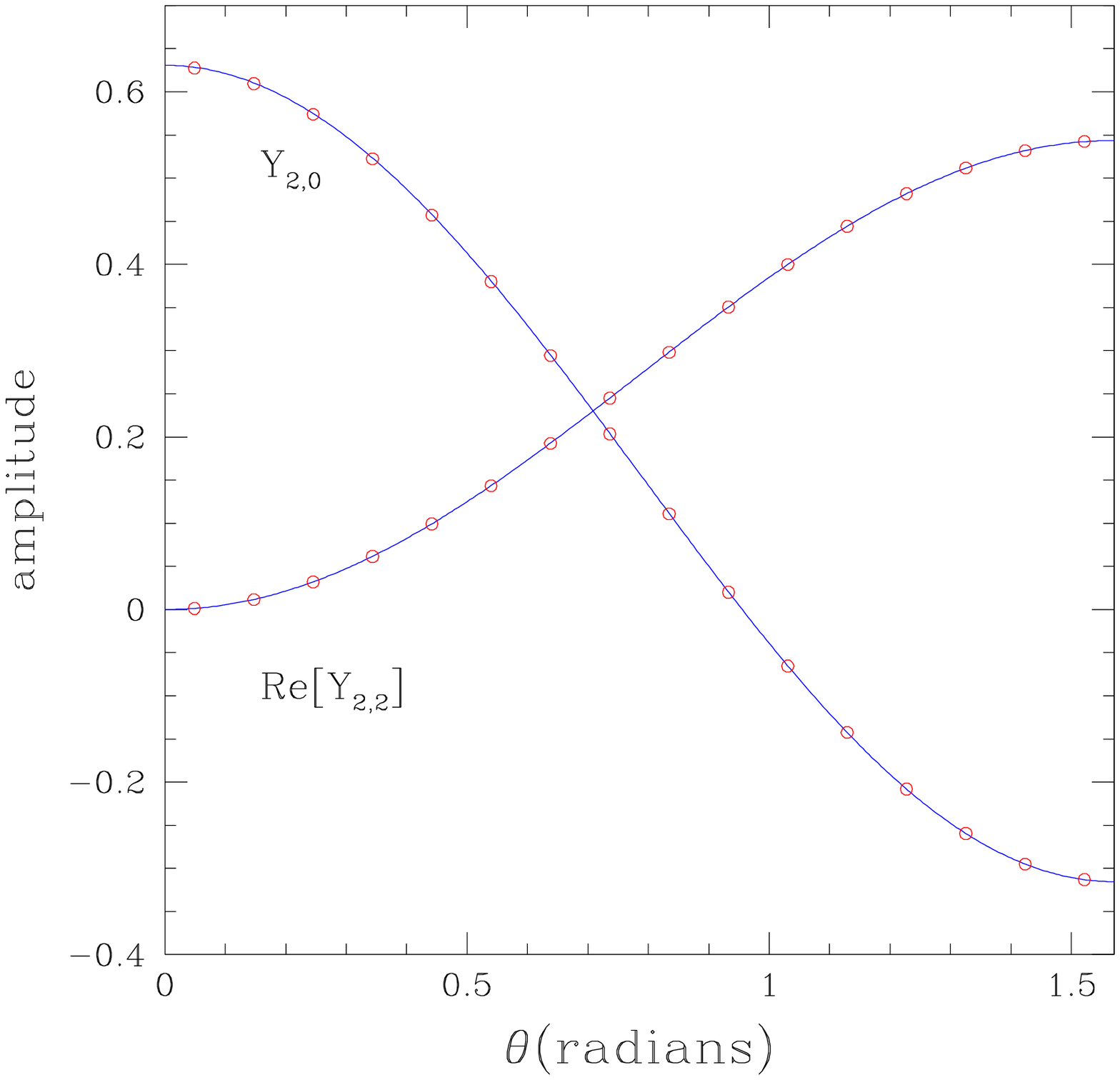}
\includegraphics[width=.45\textwidth]{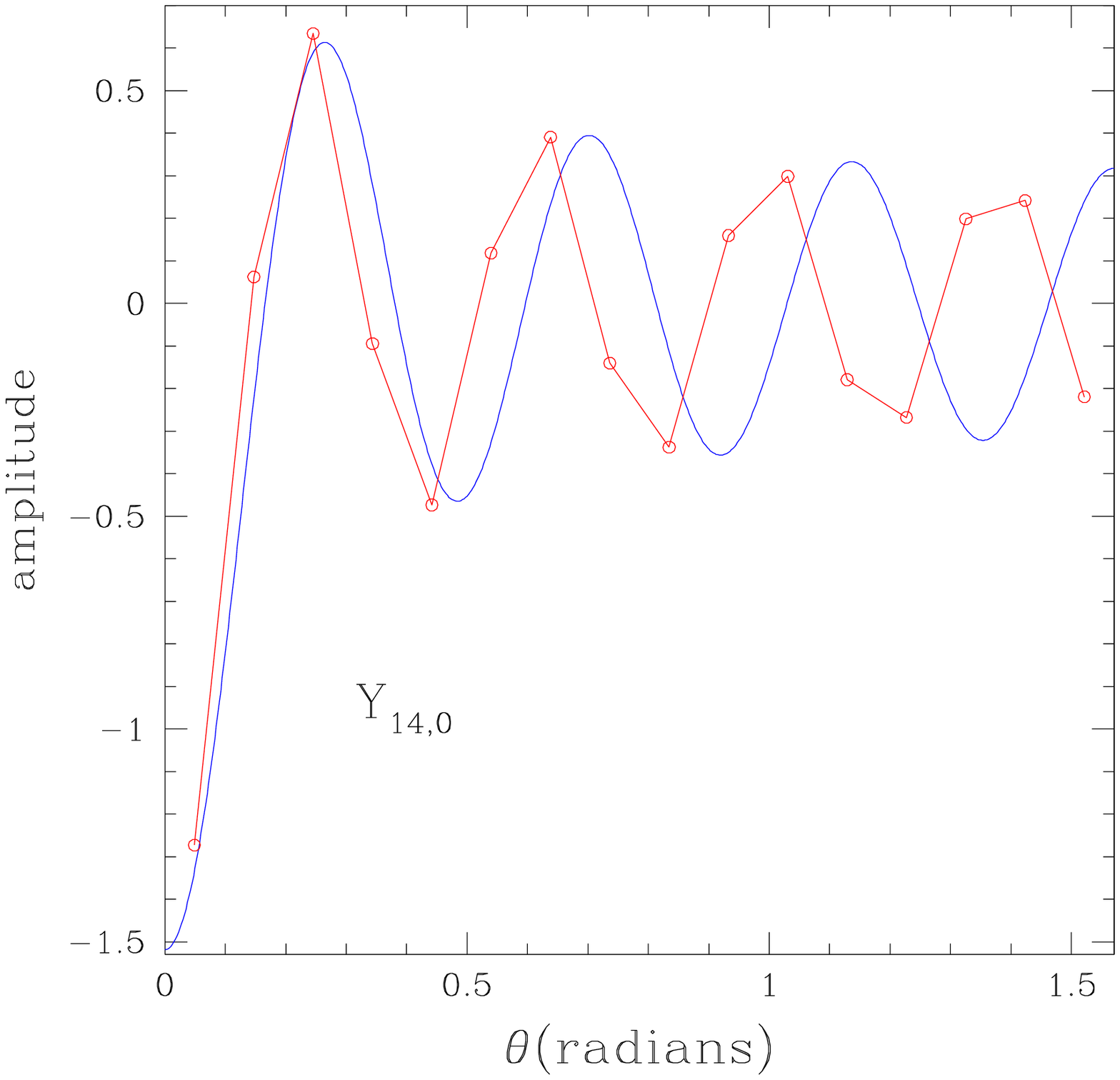}
\caption{The eigenvectors for a $16\times32
$ grid compared to the corresponding continuum eigenfunctions,
the spherical harmonics. The continuous curves show 
the spherical harmonics; the data points are 
the components of the eigenvectors.
\label{fig:roseshow2}}
\end{figure}

The $i,j$ component of the eigenvector will have the form $Y^{(k)}
_{ij} $ which should be a good approximation to some $Y_{\ell
m}(\Theta_i ,\Phi_j )$, i.e.\,, to some continuum spherical harmonic
evaluated at  grid points. (In practice we work only with real
eigenvectors that are approximations to normalized real and imaginary
parts of the grid-evaluated spherical harmonics.)
In Fig.~\ref{fig:roseshow2} continuum spherical harmonics are compared
to the eigenvectors found for a grid with $n_\Theta\times n_\Phi
$=$16\times32$ on an angular domain $0\leq\Theta\leq\pi/2 $,
$0\leq\Phi\leq\pi$.  As might be expected, the agreement between
eigenvector and continuum function is quite good when the scale for
change of the continuum function is long compared to the spacing 
betwen grid points. 

\begin{figure}[ht] 
\includegraphics[width=.45\textwidth]{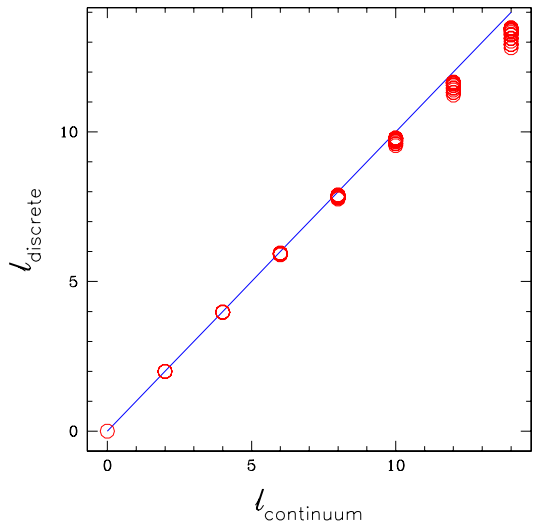}
\caption{The $\ell$ values of the discrete angular Laplacian on a $16\times32$
grid compared with the integer $\ell$ values of the continuum angular 
Laplacian. The eigenvectors of the discrete angular Laplacian
are not degenerate, so a cluster of several $\ell$ values of the eigenspectral
method corresponds to a single $\ell$ value of the continuum problem.
\label{fig:roseshow}}
\end{figure}

The eigenvalues found for the discrete and continuum angular
Laplacians are in good agreement for small eigenvalues.  For the
discrete problem we define an effective multipole index $\ell$ in the
obvious way, by setting $-\ell(\ell+1)$ equal to the eigenvalue for
each eigenvector.  A comparison is given in Fig.~\ref{fig:roseshow} of
the integer continuum values of $\ell$ and those found for a
$16\times32$ grid on the region $0\leq\Theta\leq\pi/2$,
$0\leq\Phi\leq\pi$.  (Unlike the spherical harmonics, the eigenvectors
are not degenerate, so there is a small range of $\ell$ values of 
the eigenspectral method corresponding to each $\ell$ of the continuum
problem.)
For the discrete operator the eigenvectors)
For our problem the other angular regions  
are related by symmetry. These symmetries also eliminate the odd values
of $\ell$ omitted from Fig.~\ref{fig:roseshow}. The figure shows that
for small $\ell$ there is good agreement between the discrete and
continuum eigenvalues. Because of this we can refer to monopole,
quadrupole, hexadecapole, \ldots{}eigenvectors without ambiguity.

In the mathematics of the grid space, two vectors $F_{ij}$
and $G_{ij}$ are taken to have an inner product
\begin{equation}\label{dotdef} 
F\cdot G\equiv \sum_{i=1}^{n_\Theta}\sum_{j=1}^{n_\Phi}
F_{ij}G_{ij}\sin(\Theta_i)
\Delta\Theta\,\Delta\Phi\,.
\end{equation}
With some care, detailed in Appendix~\ref{app:ESMdetails}, we can
construct the FDM angular Laplacian to be self adjoint with respect to
the inner product in Eq.~(\ref{dotdef}). This guarantees that 
the eigenvectors can be chosen to be orthogonal. We complete the 
analogy to the spherical harmonics by choosing the eigenvectors 
to be normalized, so that we have 
\begin{equation}\label{orthonorm} 
\sum_{i=1}^{n_\Theta}\sum_{j=1}^{n_\Phi}
Y^{(k)} _{ij} Y^{(k')} _{ij} 
\sin(\Theta_i)\Delta\Theta\,\Delta\Phi=\delta_{kk'}\,.
\end{equation}
(This normalization has been used for the eigenvectors shown 
in Fig.~\ref{fig:roseshow2}.)
With these definitions we can now write a multipole expansion as
\begin{equation}\label{eigensum} 
\Psi(\chi,\Theta_i,\Phi_j)=\sum_k a^{(k)}(\chi)
Y^{(k)} _{ij}\,,
\end{equation}
with 
\begin{equation}
a^{(k)}(\chi)=\sum_{i=1}^{n_\Theta}\sum_{j=1}^{n_\Phi}
\Psi(\chi,\Theta_i,\Phi_j)
Y^{(k)} _{ij}\sin(\Theta_i)\Delta\Theta\Delta\Phi\,,
\end{equation}

The multipole filtering that was the motivation for the introduction
of the spectral decomposition is implemented simply by limiting the
terms included in the sum in Eq.~(\ref{eigensum}). Rather than include
all eigenvectors, only those with $\ell<\ell_{\rm max} $ are included.
Since the discrete $\ell$s are never larger than the continuum
$\ell$s, a choice $\ell_{\rm max}=5$ means that the monopole,
quadrupole and octupole terms, with $\ell\approx0,2,4$ are
included. The effect of the eigenspectral method and multipole
filtering are the suppression of the large FDM boundary-related errors, as
illustrated in Fig.~\ref{fig:rezcav}.

In this method the $k$ summation in Eq.~(\ref{eigensum}) stops at some
maximum value governed by $\ell_{\rm max}$, and the equations to be
solved are the following modifications of
Eqs.~(\ref{radialODE}):
\begin{equation}\label{eigenspecODE} 
\sum_{k}\alpha_{k'k}\frac{d^2a^{(k)}
(\chi)}{d\chi^2} +\beta_{k'k}a^{(k)}(\chi)
+\gamma_{kk'}\frac{d a^{(k)}(\chi)}{d\chi}=S_{k'}\,.
\end{equation}
In place of 
Eq.~(\ref{alphbetgam})
the coefficients in this sum are now evaluated from
\begin{eqnarray}
\alpha_{k'k}&=&Y^{(k')}\cdot A_{\chi\chi}Y^{(k)}\\
\beta_{k'k}&=&
Y^{(k')}\cdot\left[
{A}_{\Theta\Theta}\frac{\partial^2Y^{(k)}}{\partial\Theta^2} 
+
{A}_{\Phi\Phi}\frac{\partial^2Y^{(k)}}{\partial\Phi^2} 
+
2{A}_{\Theta\Phi}\frac{\partial^2Y^{(k)}}{\partial\Phi\partial\Theta} 
+
{B}_{\Theta}\frac{\partial Y^{(k)}}{\partial\Theta} 
+
{B}_{\Phi}\frac{\partial Y^{(k)}}{\partial\Phi}
\right]\\ 
\gamma_{k'k}&=&Y^{(k')}\cdot
\left[
2{A}_{\chi\Theta}\frac{\partial Y^{(k)}}{\partial\Theta} 
+2{A}_{\chi\Phi}\frac{\partial Y^{(k)}}{\partial\Phi} 
+{B}_{\chi}Y_{\ell m}
\right]\label{eigenabc}\,,
\end{eqnarray}
where it is understood that the angular derivatives are computed
by finite differencing.
In the effective source term,
\begin{equation}\label{eigensource} 
S_{k'}\equiv -\lambda Y^{(k')}\cdot F\left(
\sum  a^{(k)}Y^{(k)}
\right)\,,
\end{equation}
only the nonlinearity appears. There is no ``true'' source term since
we solve only outside the source and introduce  the properties of the source
through boundary conditions.

Our method is clearly spectral in flavor, but it is worth pointing out
explicitly that this method is not a spectral method according to the
meaning usually given to that term in numerical analysis. If it were a
spectral, or pseudospectral (collocation) method, then angular
derivatives in the field equations and boundary conditions would be
taken using properties of the spectral functions. (If the
decomposition were done into continuum spherical harmonics, for
example, a spectral method would evaluate
$\partial\Psi/\partial\Theta$ by using relations among the spherical
harmonics.) In our method, angular derivatives are taken by finite
differencing, not by relations among the eigenvectors and their
angular derivatives. We could, in principle, convert our method to one
that meets the ``spectral method'' (actually pseudospectral)
definition. We could use finite differencing to compute, once and for
all, relations among the eigenvectors and their derivatives. These
relations could then be used to replace any derivative by a linear
combination of eigenvectors. We have, however, not explored this
approach.

Some comment must be made about a subtle but fundamental point in our
spectral method. For a given $\chi<a $, the angular specification
$\Theta=\pi/2$ refers to a single point on the $\widetilde{Z}$ axis;
the value of $\Phi$ is irrelevant. On the other hand, the function
$Y_{\ell m}(\pi/2,\Phi)$, for even $\ell $ and $m\neq0$ is, in
general, not a single value. There are, then, terms in
Eq.~(\ref{expansion}) that in principle are multivalued at $\chi<a
$,$\Theta=\pi/2$. We can, of course, delete the value $\Theta=\pi/2$
from our grid. (And we, in fact, delete this value for several
reasons, such as the requirement that the FDM Laplacian be
self-adjoint; see Appendix~\ref{app:ESMdetails}.)  We still have the
problem that the variation of these awkward terms diverges as
$\Delta\Theta\rightarrow0$ and the grid converges to the continuum.
In principle, for any $\Delta\Theta$ the summation in
Eq.~(\ref{expansion}) at any grid point will approach (in the mean)
the correct answer if we include enough multipoles.

In practice, we include very few multipoles. We must therefore ask
whether the summation will give a highly inaccurate answer in the
region of the $\chi<a$ grid near $\Theta=\pi/2$.  We avoid this
problem by choosing source structures that are symmetric about the
$\widetilde{Z}$ axis. This means that at some inner boundary
$\chi_{\rm min}$ we set the nonaxisymmetric $a^{(k)}$ to zero. The
radial equations, the FDM eigenspectral versions of
Eq.~(\ref{radialODE}), do mix the $a^{(k)}$, so the nonaxisymmetric
$a^{(k)}$ will be generated. But the mixing of the multipoles is small
until $\chi$ is on the order of $a$.  As a consequence, the
nonaxisymmetric $a^{(k)}$ can play their needed role in the wave
region without generating large errors in the near-source region.

\begin{figure}[ht] 
\includegraphics[width=.45\textwidth]{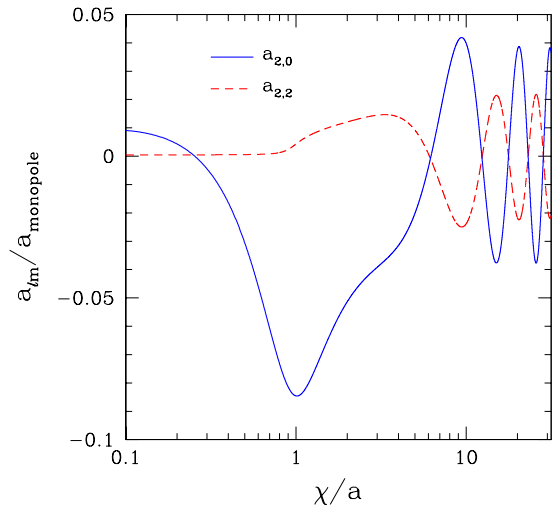}
\caption{    
The $\chi$ dependence of the eigenspectral mode coefficients. 
The solid curve shows the coefficient of the mode $Y_{20}$ 
that is symmetric about the $\widetilde{Z}$ axis; the 
dashed curve shows the real part of $Y_{22}$. In both cases
the plot shows the coefficients divided by the monopole coefficient. 
\label{fig:nonaxi}}
\end{figure}

This behavior of the coefficients is illustrated in
Fig.~\ref{fig:nonaxi} for an outgoing linear wave. The solid curve
shows, as a function of $\chi $, the eigenspectral coefficient
$a^{(k)} $ corresponding to the $\ell=2 $ mode that is symmetric about
the source axis $\widetilde{Z} $, that is, the mode corresponding to
$Y_{20}(\Theta,\Phi)$; the dashed curve shows the eigenspectral mode
corresponding to the real part of $Y_{22}(\Theta,\Phi)$. In both
cases, the value of the coefficient is divided by the value of the
monopole coefficient to give a better idea of the relative importance
of the mode in determining the overall angular behavior.  The
$Y_{20}(\Theta,\Phi)$ mode, which does not involve multivalued
behavior on the $\widetilde{Z}$ axis has a nonnegligible coefficient
at small $\chi$. By contrast the $Y_{22}(\Theta,\Phi)$, which is
multivalued, has a very small coefficient, one that is two orders of
magnitude smaller than the monopole, up to $\chi\approx1$.  For
larger $\chi$ this mode gets ``turned on,'' as it must, since it is
part of the radiation.

\section{Models and methods}\label{sec:modmeth} 

\subsection*{Nonlinear scalar models}
The model problem of Paper I, in the original comoving spherical 
coordinate system is
\begin{equation}\label{ourprob} 
{\cal L}(\Psi)=\sigma_{\rm eff}[\Psi]\,,
\end{equation}
with 
${\cal L}$ taken to be
\begin{equation}\label{ourL} 
{\cal L}\equiv 
\frac{1}{r^2}\frac{\partial}{\partial r}
\left(r^2\frac{\partial}{\partial r}\right)
+\frac{1}{r^2\sin\theta}\frac{\partial}{\partial\theta}
\left(\sin\theta\frac{\partial}{\partial\theta}\right)
+\left[\frac{1}{r^2\sin^2\theta}
-\Omega^2
\right]\,\frac{\partial^2}{\partial\varphi^2}\,,
\end{equation}
and with the effective source terms 
\begin{equation}\label{oursigma} 
\sigma_{\rm eff}[\Psi]=
\mbox{point source}-\lambda F\,.
\end{equation}
In Paper I an explicit delta function term was used in $\sigma_{\rm
eff}$ to represent the point source. Here we compute only outside the
source and include source effects by the inner boundary conditions
described below.

Our choice of the nonlinearity function $F$ is
\begin{equation}\label{modelF} 
F=\frac{1}{a^2}
\frac{\Psi^5}{\Psi_0^4+\Psi^4}\,,
\end{equation}
in which $\Psi_0$ is an adjustable parameter that we set to 0.15 or
0.01 in the numerical results to be reported.  As detailed in Paper I,
this choice of $F$ allows us to make useful estimates of the
action of the nonlinearity. We briefly review this feature here.

We define $R$ to mean distance from a source point, and we identify
$R_{\rm lin}$ as the 
characteristic distance separating the $|\Psi|\gg|\Psi_0|$
near-source  nonlinear region, and the $|\Psi|\ll|\Psi_0|$
distant region in which nonlinear effects are negligible.
In the nonlinear region near a source of strength $Q/a$
the solution  approximately has the Yukawa form
\begin{equation}\label{yukawa} 
\Psi\approx \frac{Q}{a}\;\frac{e^{-\sqrt{-\lambda\;}R/a}
}{4\pi R/a}\quad\quad\quad\mbox{\rm near source pt}\,.
\end{equation}
We can estimate $R_{\rm lin}$ by taking it to be the value of $R$ at
which the expression in Eq.~(\ref{yukawa}) is equal to $\Psi_0$:
\begin{equation}\label{estimate1} 
\frac{Q}{a}\;\frac{e^{-\sqrt{-\lambda\;}R_{\rm lin}/a
}}{4\pi R_{\rm lin}/a}=\Psi_0\,.
\end{equation}
If $R_{\rm lin}$ is significantly less than $a$, which it is for 
most of the models we consider, then we can approximate $\Psi
$ as having the Yukawa form in Eq.~(\ref{yukawa}) out to 
$R_{\rm lin}$. For $R>R_{\rm lin}$ the linear Coulombic
form should apply. We can therefore view 
$\exp{(-\sqrt{-\lambda\;}R_{\rm lin}/a)}$
as a factor by which the strength of the source is reduced. 
Since the waves are generated at distances from the source
much greater than 
$R_{\rm lin}$, the wave amplitude as well as the monopole
moment of the source should be reduced by this factor. 
We saw in Paper I that these estimates were in reasonably
good agreement with the results of computation, good enough
to give confidence of the fundamental correctness of the picture 
on which the estimate is based. We therefore use this picture
in the present paper in interpreting some of the computational
results.

\subsection*{Boundary conditions}

In Paper I the source was taken to be  two unit point
charges moving at radius $a
$
\begin{equation}\label{ptsource} 
S=\gamma^{-1}
\frac{\delta(r-a)}{a^2}
\;\delta(\theta-\pi/2)\left[\delta(\varphi)+\delta(\varphi-\pi)
\right]\,.
\end{equation}
Here $\gamma$ is the Lorentz factor $1/\sqrt{1-a^2\Omega^2\;}$.  This
factor is necessary if the source is to correspond to points of unit
strength as measured in a frame comoving with the source points. (This
factor was inadvertently omitted from the source in Paper I. In that
paper only the case $a\Omega =0.3$ was studied, so we may consider the
point sources in Paper I not to have been unit scalar charges, but
source points with charges $\gamma=1.048$.)  In the present paper we
specify inner boundary conditions on some surface $\chi_{\rm min}$
rather than an explicit source term as in Paper I. 
Our standard choice for
the inner boundary conditions will be those that correspond to the 
point sources of Eq.~(\ref{ptsource}). For this choice of source
and for $\chi_{\rm min}\ll a$ and $\ll (-\lambda)^{-1/4}a$, we can 
use an approximation for a single source point.

In notation appropriate to the 3D case we have
\begin{equation}
R^2=\tilde{Z}^2+\gamma^2 (\tilde{X}-\beta t)^2
+\tilde{Y}^2
\end{equation}
Now we use the transformations  
of Eq.~(\ref{zof}) --(\ref{yof})
to get
\begin{equation}
R^2=\frac{\chi^4}{4}
+\left(\gamma^2-1
\right)\left(\tilde{X}-\beta t\right)^2
+{\cal O}(\chi^6)
=\frac{\chi^4}{4}\left[1
+\left(\gamma^2-1\right)\sin^22\Theta\cos^2\Phi
\right]
\end{equation}

For a unit strength source at position 1, the field near position 1,
due to source 1, should be
\begin{equation}\label{eqn25} 
\psi=\frac{1}{4\pi}\frac{1}{R}\approx\frac{1}{4\pi}\frac{2}{\chi^2
}\,\frac{1}{\sqrt{1+\left(\gamma^2-1\right)
\sin^22\Theta\cos^2\Phi
\;}}
\end{equation}

The outer boundary condition in our computation is based on the the
Sommerfeld condition in Eq.~(\ref{FDMoutbcapprox}); the ingoing
condition is identical except for a change of sign. These radiative
boundary conditions 
should be applied only to the radiative part of the
wave. This is done by applying the conditions to the sum on the
right side of Eq.~(\ref{eigensum}) with the monopole mode omitted.
The multipole components of this outer boundary condtion 
are then projected out.
The monopole moment, of course, is nonradiative. Since it falls off 
at large distances as $1/\chi$, the outer boundary condition is taken to be 
\begin{equation}\label{outerbc} 
\left(\frac{da^{(0)}}{d\chi}
+\frac{a^{(0)}}{\chi}\right)_{\chi_{\rm max}}=0\,.
\end{equation}

\subsection*{Extraction of an outgoing approximation}
In Paper I, we explained how to extract a good approximation 
of the outgoing solution from the computed  standing-wave solution.
That explanation started with the solution of the linearized problem
\begin{equation}\label{extra1} 
\Psi_{\rm stndcomp}
=
\sum_{\mbox{even $\ell$}}
\ \
\sum_{m=0,\pm2,\pm4..}
\alpha_{\ell m}(r)Y_{\ell m}(\theta,\varphi)\,.
\end{equation}
We keep that notation here, but understand that (i) the role of the
continuum spherical harmonics is played by appropriate linear
combinations the eigenvectors, that (ii) the role of the
coefficients $a_{\ell m}$ is played by appropriate linear combinations of
the coefficients $a^{(k)}(\chi)$, and that (iii) the summations
only extend up to $\ell_{\rm max}$.

As in Paper I, this form of the computed standing-wave solution is
compared with a general homogeneous linear ($\lambda=0$) standing-wave
(equal magnitude in- and outgoing waves) solution of, with the
symmetry of two equal and opposite sources:
\begin{equation}
\Psi_{\rm stndlin}=
\sum_{\mbox{even $\ell,m$}}
\ Y_{\ell m}(\theta,\varphi)\,\left[
\textstyle{\frac{1}{2}}\;
C_{\ell m}h^{(1)}
_{\ell}(m\Omega r)
+\textstyle{\frac{1}{2}}\;C^*
_{\ell m}h^{(2)}
_{\ell}(m\Omega r)
\right]\,.
\end{equation}  
A fitting, in the weak-field zone, of this form of the
standing-wave multipole to the computed function $\alpha_{\ell m}(r)$
gives the value of $C_{\ell m}$.

By viewing the linear solution as half-ingoing and half-outgoing
we define the extracted outgoing solution to be
\begin{equation}\label{extractfirst} 
\Psi_{\rm exout}=
\sum_{\mbox{even $\ell$}}
\ \
\sum_{m=0,\pm2,\pm4..}
Y_{\ell m}(\theta,\varphi)\,
C_{\ell m}h^{(1)}
_{\ell}(m\Omega r)\ .
\end{equation}
Since this extracted solution was fitted to the computed solution
assuming only that linearity applied, it will be a good approximation
except in the strong-field region. In the problems of interest, the
strong-fields should be confined to a region near the sources. In
those regions, small compared to a wavelength, the field will
essentially be that of a static source, and will be insensitive to the
distant radiative boundary conditions. As 
explained in Paper I, the solutions in this region will be essentially
the same for the ingoing, outgoing, and standing-wave problem.
In this inner region then, we take our extracted outgoing solution
simply to be the computed standing-wave solution, so that
\begin{equation}
\Psi_{\rm exout}=\left\{
\begin{array}{ll}
\sum Y_{\ell m} C_{\ell m}h^{(1)}_{\ell}\ \ &\mbox{weak field outer region}
\\
\Psi_{\rm stndcomp}
&\mbox{strong field inner region}
\end{array}
\right.\ .
\end{equation}

The transition between a strong field inner region and weak field
outer region can be considered to occur in some range of $\chi$. The
maximum $\chi $ in this range must be small compared to the wavelength
$1/\Omega$, and the minimum $\chi$ must correspond to a distance from
the source larger than our estimate of $R_{\rm lin}$. [See
Eq.~(\ref{estimate1})]. For distances $R$ from the source that are of
order $a $ or less, $\chi\approx\sqrt{2aR\;}$ so the the minimum
$\chi$ in the transition region should be larger than
$\chi\approx\sqrt{2aR_{\rm lin}\;}$.

In order for the extracted solution to be smooth at this
boundary, we construct our extracted 
solution by using a blending of the strong-field inner 
solution and the weak-field outer solution over a range
from 
$\chi_{\rm low}$  to $\chi_{\rm high}$. 
In this range we take
\begin{equation}\label{extracteq} 
\Psi_{\rm exout}=\beta(\chi)\;\sum Y_{\ell m} C_{\ell m}h^{(1)}_{\ell}
+[1-\beta(\chi)]\;\Psi_{\rm stndcomp}\,.
\end{equation}
Here
\begin{equation}\label{extractlast} 
\beta(\chi)\equiv  
3\left[\frac{\chi-\chi_{\rm low}}{\chi_{\rm high}-\chi_{\rm low}}\right]^2
-2\left[\frac{\chi-\chi_{\rm low}}{\chi_{\rm high}-\chi_{\rm low}}\right]^3
\,,
\end{equation}
so that $\beta(\chi)$ goes from 0 at $\chi=\chi_{\rm low}$ to unity at
$\chi=\chi_{\rm high}$ and has a vanishing $\chi$-derivative at both
ends.  In principle we should choose $\chi=\chi_{\rm high}$ to depend
on the location of the wave zone, and hence on $\Omega$, and in
principle we should choose 
$\chi=\chi_{\rm low}$ to depend on the nature of the nonlinearity, and 
hence on $\lambda$ and $\Psi_0$. In practice we have found 
it to be adequate to choose 
$\chi_{\rm high}=2a$ and $\chi_{\rm high}=3a$ 
for all models.

\subsection*{Nonlinear iteration}
The computational problem of finding a solution $\Psi $ consists of
finding a set of coefficients $a^{(k)}(\chi)$ that satisfy the field
equation Eqs.~(\ref{eigenspecODE}) along with the inner and outer
boundary conditions.  The operations on the left hand side of
Eqs.~(\ref{eigenspecODE}) are linear on the $a^{(k)}(\chi)$, as are
the boundary conditions, so the problem of finding the
$a^{(k)}(\chi)$ can be written as
\begin{equation}
\sum_k
{\cal L}_{k'k}
a^{(k)}
={\cal F}_{k'}(\{a^{(p)}\})\,,
\end{equation}
where ${\cal L}_{k'k}
$ is a linear differential operator on the
$a^{(k)}(\chi)$, and where ${\cal F}_{k'}
(\{a^{(p)}\}$, containing the
nonlinearity in the model, is nonlinear in the $a^{(k)}(\chi)$.

For different boundary conditions (outgoing or ingoing) the linear
operator ${\cal L}_{k'k}$ has different forms, but in either form we
can invert to get the outgoing or ingoing Green functions 
${\cal L}^{-1,\rm out}_{k'k}
$ and ${\cal L}^{-1,\rm in}_{k'k}$. In principle we can then 
find solutions by direct iteration
\begin{displaymath}
a^{(k')}_{n+1,\rm out}=\sum_k
{\cal L}^{-1,\rm out}_{k'k}
\left(
{\cal F}_{k'}(\{a^{(p)}_{n,\rm out}\})
\right)
\quad\quad\quad
a^{(k')}_{n+1,\rm in}=\sum_k
{\cal L}^{-1,\rm in}_{k'k}
\left(
{\cal F}_{k'}(\{a^{(p)
}_{n,\rm in}\})
\right)
\end{displaymath}
\begin{equation}\label{it4outin} 
a^{(k')}_{n+1,\rm stnd}=\sum_k
\textstyle{\frac{1}{2}}\left\{
{\cal L}^{-1,\rm out}_{k'k}+{\cal L}^{-1,\rm out}_{k'k}
\right\}
\left(
{\cal F}_{k'}(\{a^{(p)}_{n,\rm stnd}\})
\right)\ .
\end{equation}

In Paper I it was pointed out that this kind of direct iteration
converges only for weak nonlinearity. More generally we use 
Newton-Raphson iteration and solve
\begin{equation}\label{newtraph}
\sum_k
\left[{\cal L}_{k'k}
-\left.\frac{\partial{\cal F}_{k'}}{\partial 
a^{(k)}
 }
\right|_{
a^{(p)}_n
}
\right]a^{(k')}_{n+1}
=
{\cal F}_{k'}\left(
\{a^{(p)}_{n}\}\right)
-\sum_{k}
\left.\frac{\partial{\cal F}_{k'}}{\partial 
a^{(k)}
 }
\right|_{
a^{(p)}_n
}
a^{(k)}_{n+1}
\end{equation}
This Newton-Raphson approach can be applied to find outgoing, ingoing
and standing-wave solutions analogous to those in
Eqs.~(\ref{it4outin}).  It has been applied with an error measure
\begin{equation}
\epsilon\equiv\sqrt{\frac{1}{k_{\rm max}n_{\chi}}{\sum_{k=1}^{k_{\rm max}
}\sum_{i=1}^{n_\chi}
\left(a^{(k)}_{n+1}(\chi_i)
-a^{(k)}_{n+1}(\chi_i)
\right)^2}\;}\,.
\end{equation}
Iteration was halted when this error measure fell below $\sim10^{-6}$.
Note that for strongly nonlinear models, convergence sometimes
required that the iteration described in Eq.~(\ref{newtraph}), had to
be somewhat modified. The last term on the right in
Eq.~(\ref{newtraph}), had to be weighted by a factor less than unity,
at least until the iteration got close to the true solution.

\section{numerical results}\label{sec:numresults}
%
\begin{figure}[ht] 
\includegraphics[width=.4\textwidth]{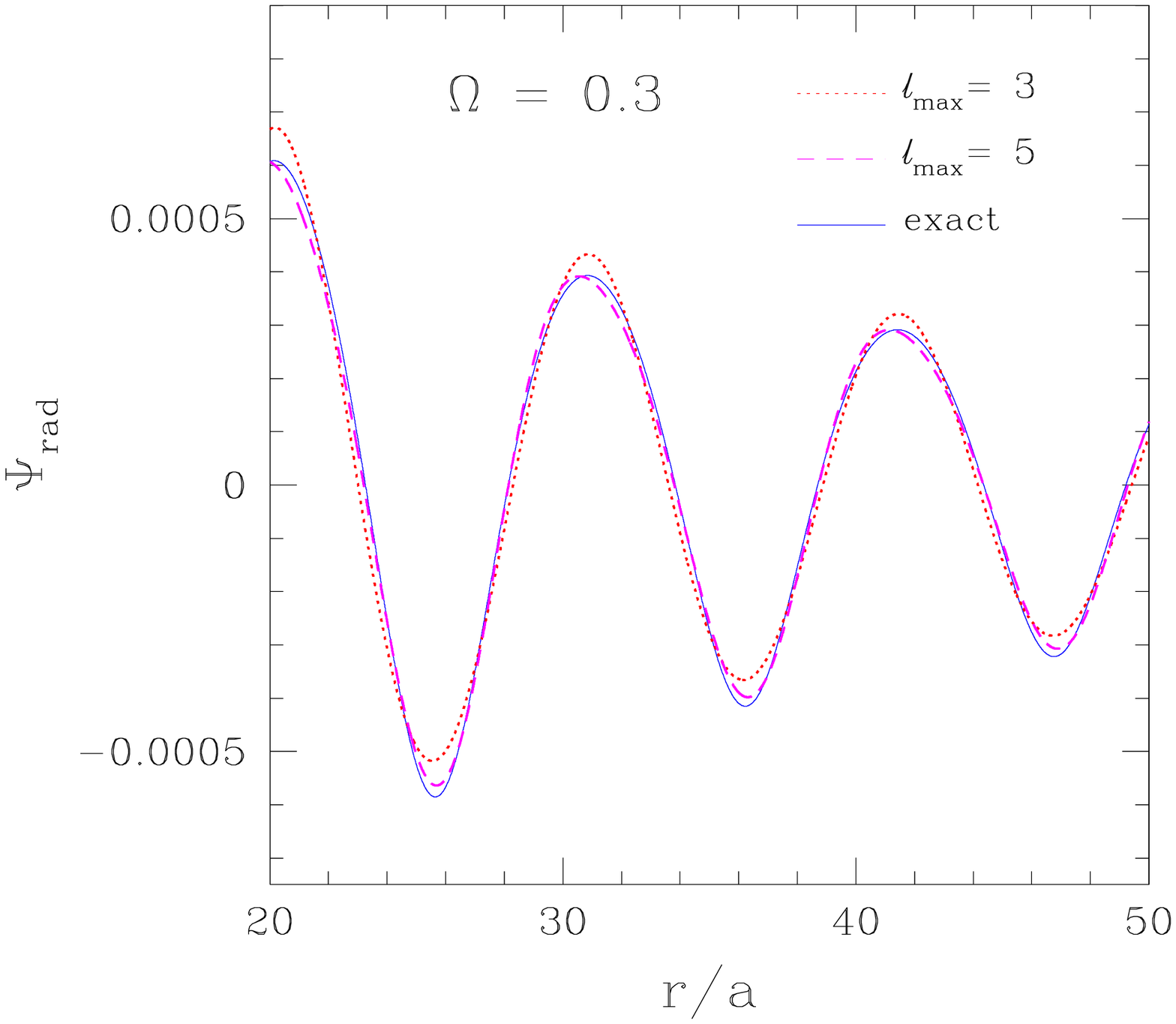}
\includegraphics[width=.4\textwidth]{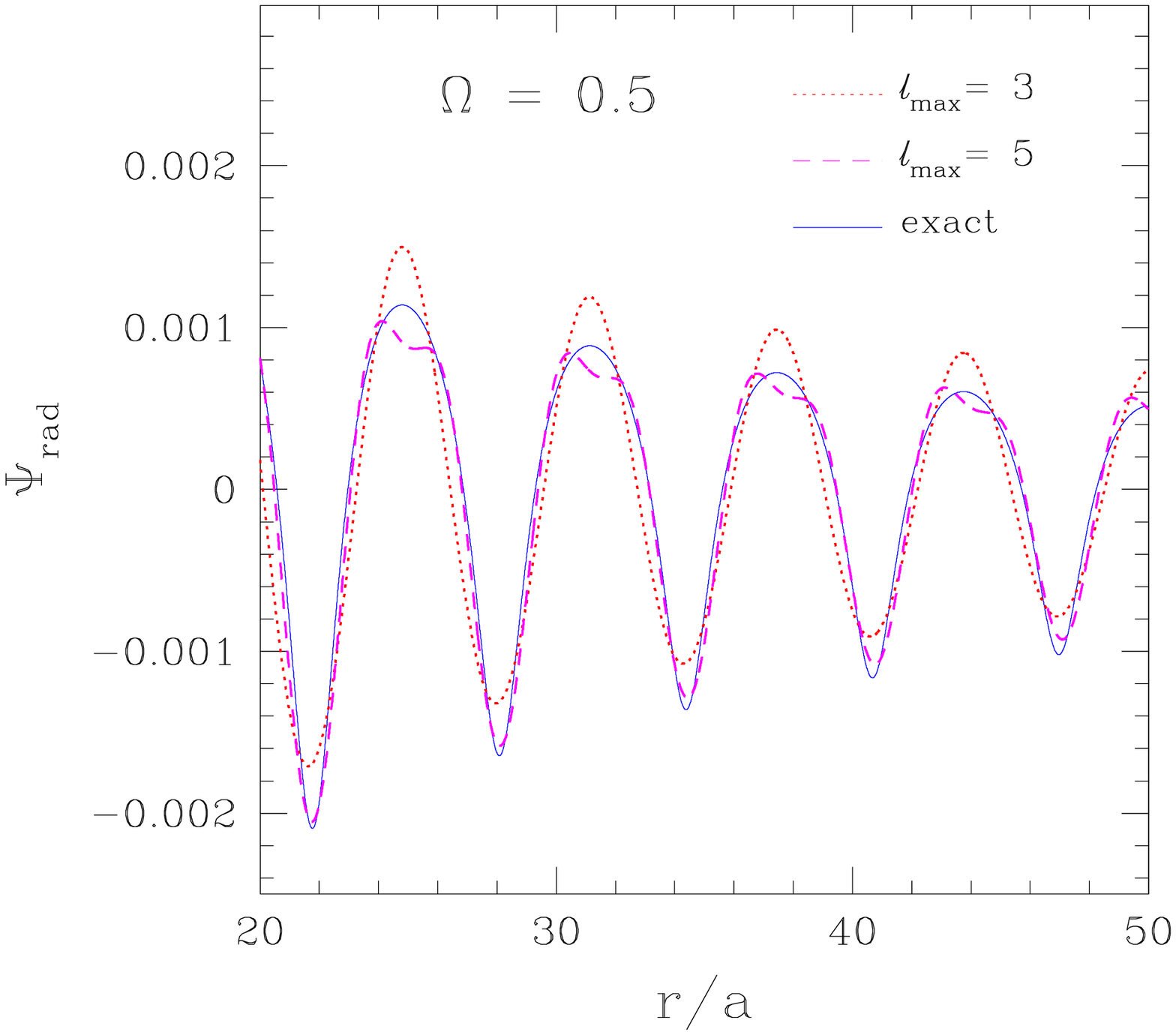}
\caption{Comparison of exact and eigenspectral linear outgoing solutions.
The solid curve shows the exact solution, in the wave region, computed from an
infinite series. The other curves show the result of computation with 
a grid with $n_{\chi}\times n_{\Theta}\times n_{\Phi}$=$12001\times16\times32$
and $\chi_{\rm min}/a=0.2$, $\chi_{\rm max}/a=75$.
Results are shown with 
$\ell_{\rm max}=3$ (monopole and quadrupole modes kept)
and $\ell_{\rm max}=5$ (monopole, quadrupole  and hexadecapole).
The results are shown at $\Theta=0$ as a function of $r$,
the distance from the center of the configuration.
\label{fig:compare}}
\end{figure}

If numerical results are to be trusted they must converge, or at least
be stable, as computational parameters (grid size, etc.) change, and
there must be evidence that the result is the correct answer to the
physical problem.  A complication in demonstrating this is that at the
same time we are making two different classes of approximations:
(i)~we use values on a grid in place of the continuum mathematics,
(ii)~we are keeping only low order multipoles. In addition, to
represent point sources we use approximations for inner boundary data
that are exact only only for $\chi_{\rm min}\rightarrow0$.  Our outer
boundary conditions in Eq.~(\ref{outerbc}) also add an error, in
principle one of order $(a/\chi_{\rm max})^2$, but we have found that
this error is negligible compared to that of our other
approximations. (Moving the boudary outward has no discernible effect
on results.) Here we present results of varying the grid resolution,
the number of multipoles kept, and the inner surface $\chi_{\rm min}$
on which inner Dirichlet data is set.
\begin{table}
\caption{Convergence for rotating linear models.
All models have $a\Omega=0.3$, $\lambda=0$, and use 
outgoing boundary conditions at
$\chi_{\rm max}=50a$. The computed monopole to source strength
index, $\gamma Q_{\rm eff}/Q$, is unity in the exact
solution. The ``two region'' computation retains all 
multipoles for $\chi<3a$.
\label{table:convlin} }
\begin{tabular}{|c|c|c|c|c|c|c|}
\hline
$n_\chi$&$n_\Theta$&$n_\Phi$&$\chi_{\rm min}/a$
&
$\ell_{\rm max}$&
$\gamma Q_{\rm eff}/Q$&two region\\
\hline

        1001 & 8 & 16 & 0.2 & 3 & 1.0116 &-\\

           2001 & 16 & 32 & 0.2 & 3 & 1.0246 &-\\

           4001 & 16 & 32 & 0.2 & 3 & 1.0275 &-\\

           8001  & 16  & 32 &  0.2 &  3 &  1.0282&-\\

          16001 & 32 & 64 & 0.2 & 3 & 1.0284 &-\\

\hline
8001 &  16  &   32  &  0.2  &   3 &  1.0282 &    -\\
 8001 &  16 &    32 &   0.2  &   5 &  1.0036 &  1.0035\\
 8001 &  16 &    32 &   0.2   &  7 &  0.9968 &  0.9966\\
 8001  & 16 &    32 &   0.2   &  9 &  0.9934 &  0.9928\\
\hline

 16001& 16 &  32 &  0.2   &  5    & 1.0037 &-\\
 16001 &16  &  32&   0.1  &   5  &1.0036    &-\\
 16001& 16  & 32  & 0.05   & 5   & 1.0032  &-\\
 16001& 16  & 32  & 0.025  & 5   & 1.0007   &-\\
\hline
\end{tabular}
\end{table}

Error is most easily measured for solutions to the linear problem
since there exists an exact series solution for comparison.  Figure
\ref{fig:compare} shows a comparison of this series solution with
computed solutions for two different source speeds $a\Omega $. The
qualitative features of these plots agree with what should be
expected: the eigenspectral/multipole filtering technique is more
accurate at lower source speed, and is more accurate when more
multipoles are allowed to pass through the ``filter.'' (Of course,
there will be a point of diminishing returns. If we let too many modes
through then we are no longer filtering, and we experience the
difficulties that plagued the FDM method for adapted coordinates.)

An obvious unwelcome feature of the results with $\ell_{\rm
max}=5$ is the 
phase error in the hexadecapole mode. This error is especially noticeable 
in the $a\Omega=0.5$ plot where it causes an artifact fine structure 
at the positive peak of the waves. We are investigating 
the source of this phase error which we suspect is a result of 
truncation error in angular differencing and/or in the computation 
of the eigenvectors. We have anecdotal evidence that the phase error
decreases as the angular grid is refined.

The reliability of the eigenspectral method for a wider variety of
linear models is presented in Table~\ref{table:convlin}.  In this
table, the measure of error is the value of $Q_{\rm eff}$, the
monopole moment computed near the outer boundary for the ``charge,''
i.e.\,, the monopole moment of the two sources each with scalar charge
$Q/a=1$.  Though the monopole moment would seem to be less interesting
than features of the radiation, we have found in essentially all
computations that the largest error is in the monopole.  For example,
the majority of the error in the computed amplitude of radiation could
be understood to be due to the error in the monopole. The error in the
ratio of radiation amplitude to monopole was several times smaller
than the raw errors in either quantity by itself.
For simplicity we use this one measurement to characterize convergence
and correctness.

To show that the computed solution is accurate it is convenient to
consider first the linear outgoing problem for two unit point charges,
since the solution for this case is known to be $Q_{\rm
eff}=1/\gamma$, where $\gamma$, the Lorentz factor, is
$1/\sqrt{1-a^2\Omega^2\;}$.
(The complete solution for $\Psi$ in this case is given as Eq.~(10) of
Paper I, though the series solution must be multiplied by $1/\gamma$
since we are now considering unit charges.)

Table~\ref{table:convlin} presents results for linear ($\lambda=0$),
rotating ($a\Omega=0.3$) models, for scalar source points with unit
charge. For all models the inner boundary conditions were those of the
small-$\chi$ point approximation given in Eq.~(\ref{eqn25}) at some
$\chi_{\rm min}$. The number of multipoles kept is specified by the
parameter $\ell_{\rm max}$. Choosing $\ell_{\rm max}=3$ means that
modes corresponding to monopole and quadrupole were kept; $\ell_{\rm
max}=5$ means that in addition the hexadecapole was kept; and so
forth.  The accuracy criterion used is the quantity $\gamma Q_{\rm
eff}/Q$, the value of which is unity in the exact solution.

The results in the table are divided into three sections. In the first
section the number of grid points $n_\chi$, $n_\Theta$, and $n_\Phi$,
was varied, while the values of $\chi_{\rm min}$ and $\ell_{\rm max}$
are kept fixed. The results show 3\% accuracy, and demonstrate that,
for the parameters of this computation there is no advantage to grid
size larger than $8001\times16\times32$.
Note that simple considerations of
truncation error do not apply, since the angular grid is not used in a
straightforward finite differencing, but rather to establish the
angular eigenvectors.

In the second section the results show that increasing $\ell_{\rm
max}$, for an adequately large grid, improves accuracy, and results
come within a fraction of a percent of the correct answer. Note that
using {\em all} the eigenvectors is equivalent to no multipole
filtering. In that case we would be simply doing finite differencing
in the multipole basis, and we would be plagued by the problems
described at the start of Sec.~\ref{sec:spectral}.  Accuracy must,
therefore, drop off when $\ell_{\rm max}$ is increased past some
optimal value. The results in the table suggest that 
the optimal value for this model and this grid size may 
be $\ell_{\rm max}\approx5$. Larger values of $\ell_{\rm max}$
are more difficult computationally, and appear to give no 
improvement in accuracy.

In the third section of the table the value of $\chi_{\rm min}$ is
decreased and, for a fairly large grid and for $\ell_{\rm max} =5$ the
results show that the errors in the inner Dirichlet data were the
dominant source of error. More important, the results show that
very high accuracy can be achieved with the eigenspectral method
using a small number of multipoles. 

In the last column of table~\ref{table:convlin} several results are
given of a ``two region'' method of computation. The motivation for
this method is that severe multipole truncation is really necessary
only in the wave zone. Closer to the source more mutlipoles can be
kept and more precise computation can be carried out.  For the results
in the last column, all multipoles up to $\ell_{\rm max}$ were kept
for grid points with $\chi<3a$; for $\chi>3a$ only the monopole and
quadrupole eigenmodes were used. The results show no increase in error
compared with  standard method, but the error in any case
is dominated by the inner boundary data, not by truncation.
\begin{table}
\caption{Convergence for rotating nonlinear models.
All models have $a\Omega=0.3$, $\lambda=-25$, $\Psi_0=0.15
$, and all use 
outgoing boundary conditions at
$\chi_{\rm max}=50a$.   The ``two region'' method retains all 
multipoles for $\chi<3a$.
\label{table:convnonlin} }
\begin{tabular}{|c|c|c|c|c|c|c|}
\hline
$n_\chi$& $n_\Theta$ & $n_\Phi$& $\chi_{\rm min}/a$ & $\ell_{\rm max}
$& $Q_{\rm eff}/Q$&two region\\
\hline  
  1001  &    8     &   16   &    0.2     &   3   &   0.3440   &-\\
  2001    &  16    &    32  &     0.2    &    3  &    0.3450 &-\\
  4001    &  16    &    32  &     0.2    &    3   &   0.3452 &-\\
  8001    &  16    &    32   &    0.2    &    3   &   0.3452 &-\\
 16001    &  32    &    64   &    0.2    &    3   &   0.3452 &-\\
                 \hline
  8001   &   16   &    32    &   0.2     &   3    &  0.3452  &  -\\     
  8001   &   16    &    32   &    0.2     &   5    &  0.3431 &     0.3424\\
  8001   &   16    &    32   &    0.2     &   7     & 0.3421  &    0.3415\\
  8001    &  16   &    32    &   0.2       & 9    &  0.3417   &   0.3410\\
            \hline
 16001  &    16   &     32   &     0.2   &    5  &    0.3230 &-\\
 16001  &    16   &     32    &    0.1    &   5   &   0.3213 &-\\
 16001   &   16    &    32     &   0.05   &   5   &   0.3198 &-\\
 16001    &  16     &   32     &   0.025   &  5    &  0.3192 &-\\
\hline
\end{tabular}
\end{table}

For nonlinear models, with $\lambda=-25$ and $\Psi_0=0.15$,
table~\ref{table:convnonlin} gives results roughly equivalent to the
linear-model results in table~\ref{table:convlin}.  Now there is no
{\it a priori} correct answer known, so we look only for convergence
of the value of the monopole moment $Q_{\rm eff}$.  (Due to the
effects of the nonlinearity, this value can be reduced well below
unity.)  The computational results in the table show few differences
from those in table~\ref{table:convlin}. Again, the answer is shown to
be stable for moderate grid size, and there is no evidence of a strong
dependence on $\ell_{\rm max}$.  The two-region computations converged
more quickly than those with a uniform multipole cutoff, and give
results in good agreement with those of the uniform cutoff standard
approach.  This two-region technique, therefore, can be considered a
computational tool that may prove useful in more difficult problems.

\begin{figure}[ht] 
\includegraphics[width=.45\textwidth]{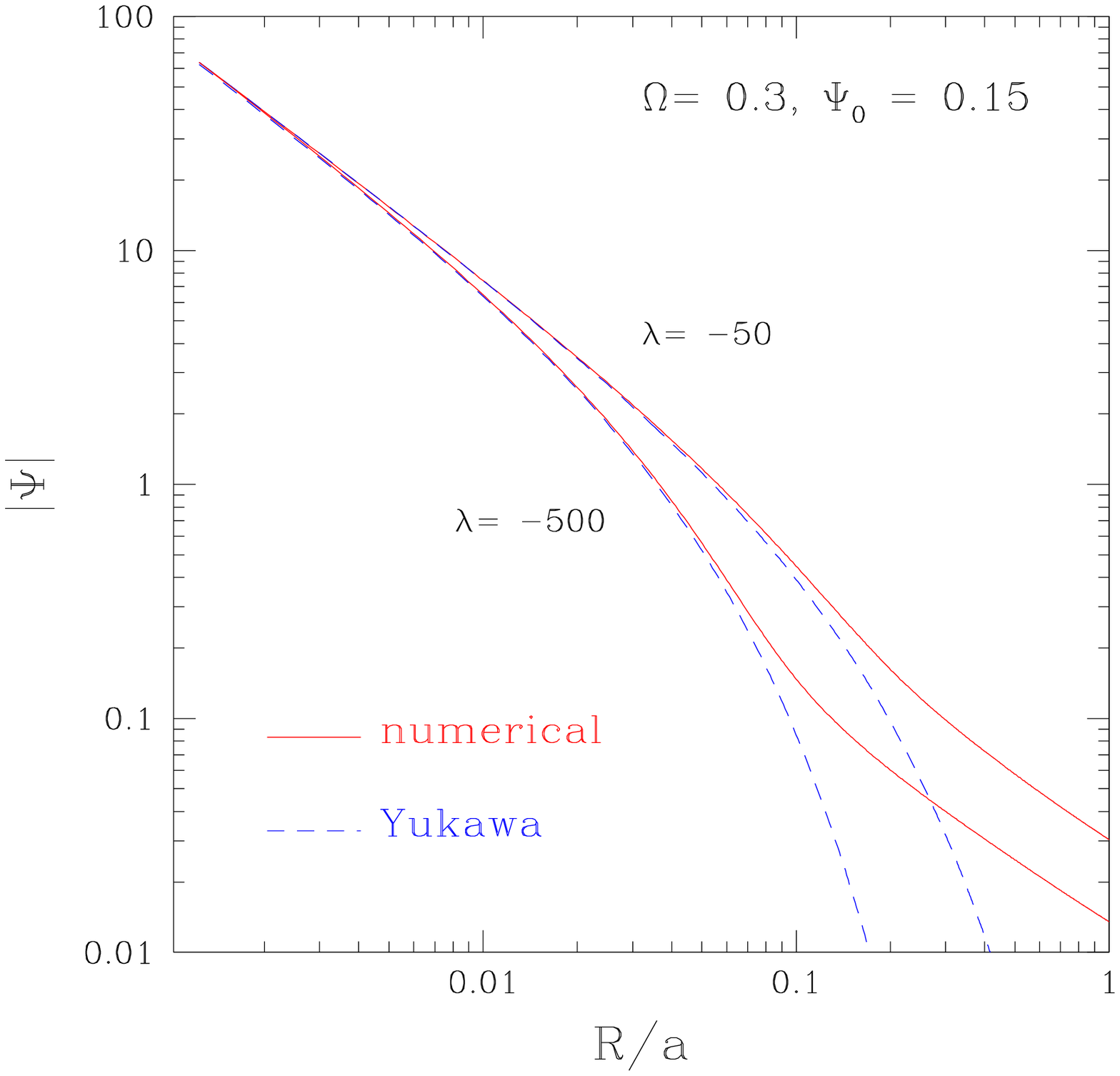}
\includegraphics[width=.45\textwidth]{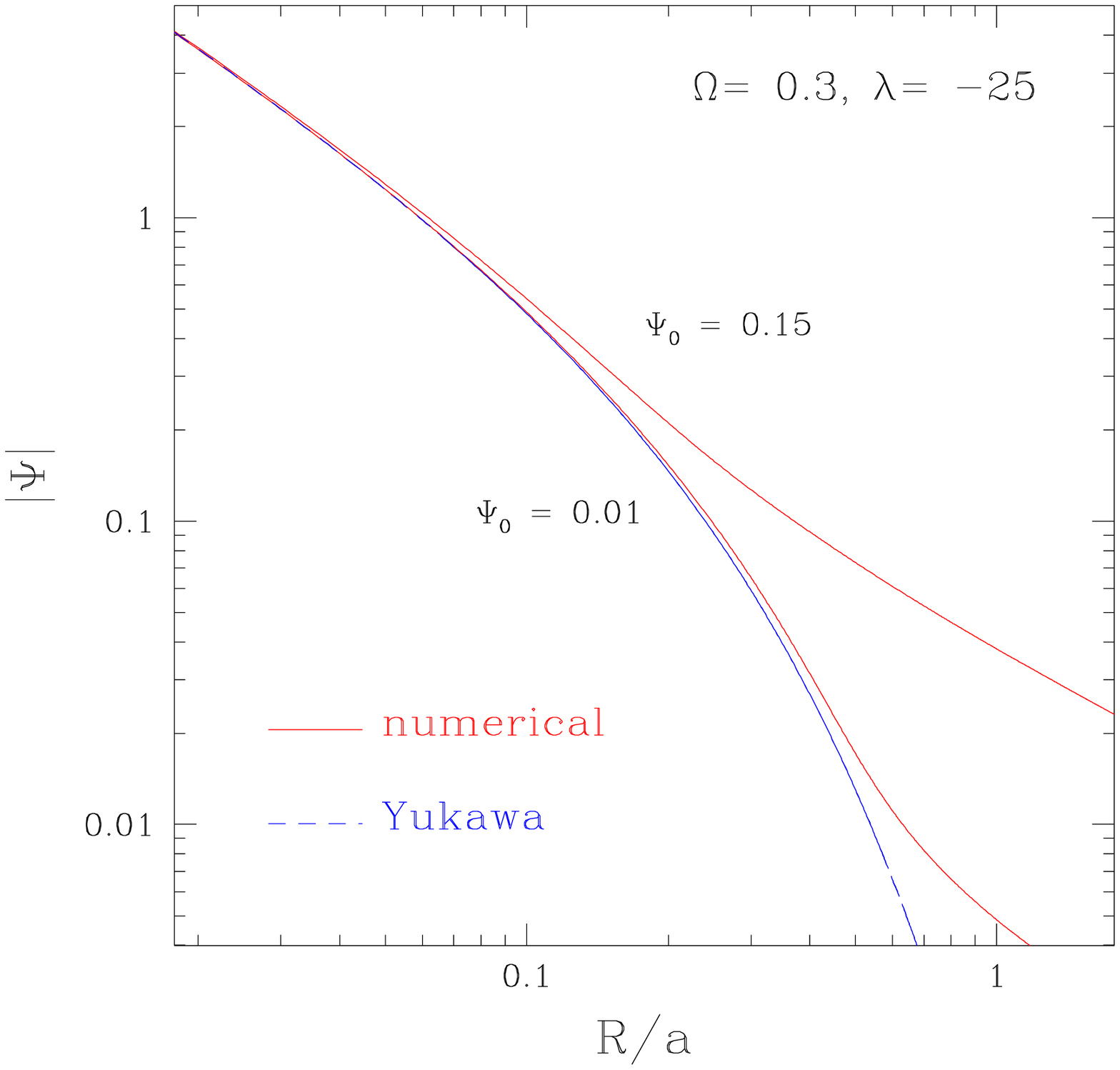}
\caption{Near-source fields for nonlinear models. 
\label{fig:yukawa}}
\end{figure}

Though there is no {\it a priori} known general solution for the nonlinear
problem, we do know one useful limit of the solution. As argued in 
Sec.~\ref{sec:modmeth},    and in Paper I, $\Psi
$ should have an approximately Yukawa form for a range of  small $\chi$.
Evidence of this in the results is presented in 
Fig.~\ref{fig:yukawa}, which
gives computed nonlinear outgoing  solutions near
the sources. The computations start with the boundary conditions
of Eq.~(\ref{eqn25}). The variable $R$ in the figure is $\widetilde{Z}-a$ along a
line through the sources, that is, the radial distance from a source.
A straight line in the log-log plots of the figure indicate that $\Psi$
is falling off approximately as $1/4\pi R$; the downard deviation from
a straight line is a manifestation of the nonlinearity. According to
the analysis following Eq.~(\ref{yukawa}), the radius $R_{\rm lin}$ at
which nonlinear effects become significant, decreases with increasing
$|\lambda|$ and with decreasing $\Psi_0$.  Results for our standard
choice $\Psi_0=0.15$ are shown on the left. The nonlinear effects
become important for $R/a$=$R_{\rm lin}/a$ on the order a few
tenths. In this case $\Psi$ is comparable to $\Psi_0$ when nonlinear
effects become important, and the Yukawa form is not distinguishable
from a $1/R$ fall off.
For more convincing evidence of the working of the nonlinearity
we
change $\Psi_0$ to 0.01. The results, shown on the right in
Fig.~\ref{fig:yukawa} for $\lambda=-25$ shows the excellent agreement
of of the computed solution to the Yukawa form in the range $R/a$=0.1
to around 0.4.

Paper I used Eq.~(\ref{estimate1})  as the basis of an estimate 
of the nonlinear solution. That estimate was applied to 
the radiation ``reduction factor,'' the factor by which the radiation
amplitude is reduced for a nonlinear model as compared with a 
model with the same parameters, but with $\lambda=0$. 
This provides us with a convenient 
comparison of the nonlinear results in Paper
I and with the present eigenspectral method. In table~\ref{table:yukawa}
we give those Paper I results again, along with eigenspectral
computations of the same models. We present additional models, 
since the standard coordinate/finite differencing method of Paper
I was
limited in the size of $\lambda$ for which Newton-Raphson runs
converged; with the eigenspectral method we can give results for much
larger values of $-\lambda$.
The last column in table~\ref{table:yukawa} gives
the computed reduction factor computed with the eigenspectral method
keeping only the monopole and quadrupole terms.  (Note: In Paper I the factor $1/\gamma
$=1/1.048 was mistakenly omitted from the delta function source. Here
we choose to treat that as a source of strength 1.048, rather than
unity. Our eigenspectral computations therefore used this enhanced
source. The estimates of $R_{\rm lin}$ were also slightly in error in
Paper I since they assumed a unit source strength. They have been
recomputed and are slightly different from the estimates presented in
Paper I.)

The agreement of the computed results with the simple estimate is
gratifying, as it was in Paper I. More important, the comparison of
the second and third columns of table~\ref{table:yukawa} shows that
the eigenspectral method with only two multipoles gives $~1\%$
agreement of the computed radiation with the very different and much
more computationally intensive finite difference method.

\begin{table}
\caption{The radiation reduction factor due to the nonlinearity.  For
all cases, $\Psi_0=0.15$, and $a\Omega=0.3$ .  The second column
refers to Eq.~(\ref{estimate1}).  The third column gives the reduction
factors presented in Paper I. The last column gives the results of the
eigenspectral computation with $\chi_{\rm min}=0.3a$, with outgoing
boundary conditions at $\chi_{\rm max}=50a$, $\ell_{\rm max}=3$, 
and a grid with
$n_\chi=8001$, $n_\Theta=16$, $n_\Phi=32$.
\label{table:yukawa} }
\begin{tabular}{|c|c|c|c|}
\hline
$\lambda$&Estimate &Paper I&Eigenspec
\\
\hline
-1&69\%&78\%&77.0\%\\
-2&62\%&68\%&68.1\%\\
-5&53\%&55\%&55.7\%\\
-10&46\%&47\%&46.4\%\\
-25&37\%&35\%&35.5\%\\
-50&25.6\%&-&28.5\%\\
-100&19.5\%&-&22.7\%\\
\hline
\end{tabular}
\end{table}
\begin{figure}[ht] 
\includegraphics[width=.5\textwidth]{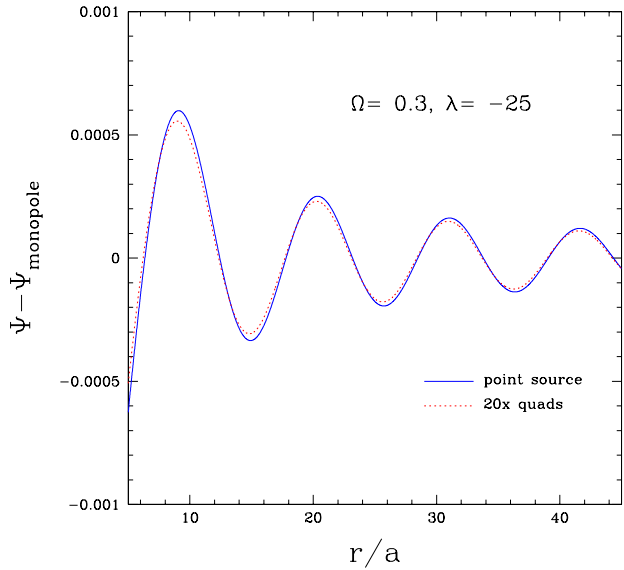}
\caption{For outgoing nonlinear waves, the sensitivity 
of the radiation to details of source multipole structure.
\label{fig:innerquad}}
\end{figure}

We have argued that the details of the higher moments of the source
are not important in determining the radiation. Some numerical
justification for this is given in Fig.~\ref{fig:innerquad}, which
shows computed results for nonlinear models with the standard
parameters. The solid curve uses the point source initial data of
Eq.~(\ref{eqn25}) as inner Dirichlet data at $\chi_{\rm min}
=0.2a$. For these inner boundary conditions the multipole moments at
$\chi_{\rm min} =0.2a$ are $a_0=-13.90$, $a_{20}=0.11045$. The
coefficient corresponding to the real part of $Y_{22}$ is 0.1920; the
coefficient corresponding to the imaginary part is zero.  We first compute the
outgoing linear solution for these inner Dirichlet data. Next we,
somewhat arbitrarily, set all the quadrupole components to -2.209,
which is 20 times the original value of $a_{20}$, and calculate the
outgoing linear radiation. The results in Fig.~\ref{fig:innerquad}
show that the effect on the radiation is of order 10\%. Some
interpretation is needed to connect this result to multipoles of
sources, especially because the effect on the radiation of a
physically plausible source quadrupole depends on the size of the
source.

If we had a source with a surface at $\chi=0.2a$ the computed result
tells us that a rather large deformation, with $|a_{2k}/a_0|\sim0.16$
will have a 10\% effect on the radiation as compared with a source
with a negligible quadrupole. With a simple argument, we can apply
this 10\% effect to sources of other size. Mathematically the
conditions at $\chi_{\rm min}=0.2a$ can be ascribed to a source with a
surface at $\chi_{\rm surf}\neq0.2a$. Quadrupole moments fall off as
$1/R^3$, where $R$ is the distance (small compared to $a$) from the
source point, and monopole moments fall off as $1/R$. The quadrupole
to monopole ratio, therefore, falls off as $1/R^2$, or $1/\chi^4$.
Since the 10\% effect corresponds to $|a_{2k}/a_0|\approx0.16$ at
$\chi_{\rm surf} =0.2a$, it also applies, to
$0.16\times(1/2)^4\approx0.01$ at $\chi_{\rm surf}=0.4a$, and to
$0.16\times(2)^4\approx2.6$ at $\chi_{\rm surf}=0.1a$.  This means
that the radiation generated is reasonably sensitive to a mild
quadrupole deformation of a source that is comparable to the size of
the binary system, but for small sources unphysically large deformations
are required to have any effect on the radiation.
(See also the related discussion of the two-dimensional case in Appendix
\ref{app:SSM2D}.)

Figure \ref{fig:extract} shows the central result of our method, the
accuracy of the outgoing approximation, for a model with $\Omega=0.3$
and $\lambda=-25$, one of the models presented in Paper
I\cite{paperI}. A measure of the strength of the nonlinearity is the
fact that the nonlinearity reduces the amplitude of the waves to 35\%
of those for $\lambda=0$ in the same model (i.e.\,, the same $\Omega$
and inner boundary data and outgoing boundary conditions). The
figure shows that the extracted solution is in remarkably good
agreement with the computed nonlinear solution in the three regions of
the extraction protocol described in Sec.~\ref{sec:modmeth}: (i)~the wave
region in which the solution is treated as a half-outgoing and
half-ingoing superposition, (ii)~the inner region in which the
outgoing solution is taken to be well approximated by the standing
wave solution since the radiative boundary conditions are irrelevant
close to the source, and (iii)~the blending region described in
Eqs.~(\ref{extracteq})--(\ref{extractlast}).

The excellent agreement
between the computed outgoing solution and the extracted approximation
should not be confused with agreement with the exact solution. As we
have seen in the comparisons of exact and computed linear solutions,
e.g.\,in Fig.~\ref{fig:compare}, the mutlipole filtering does entail
an inaccuracy of a percent or so.
Figure~\ref{fig:extract}, then, is not a demonstration of the accuracy
of the eigenspectral method, but rather a powerful statement about
effective linearity, the accuracy of the process of extracting an
outgoing approximation from a standing wave solution.
\begin{figure}[ht] 
\includegraphics[width=.4\textwidth]{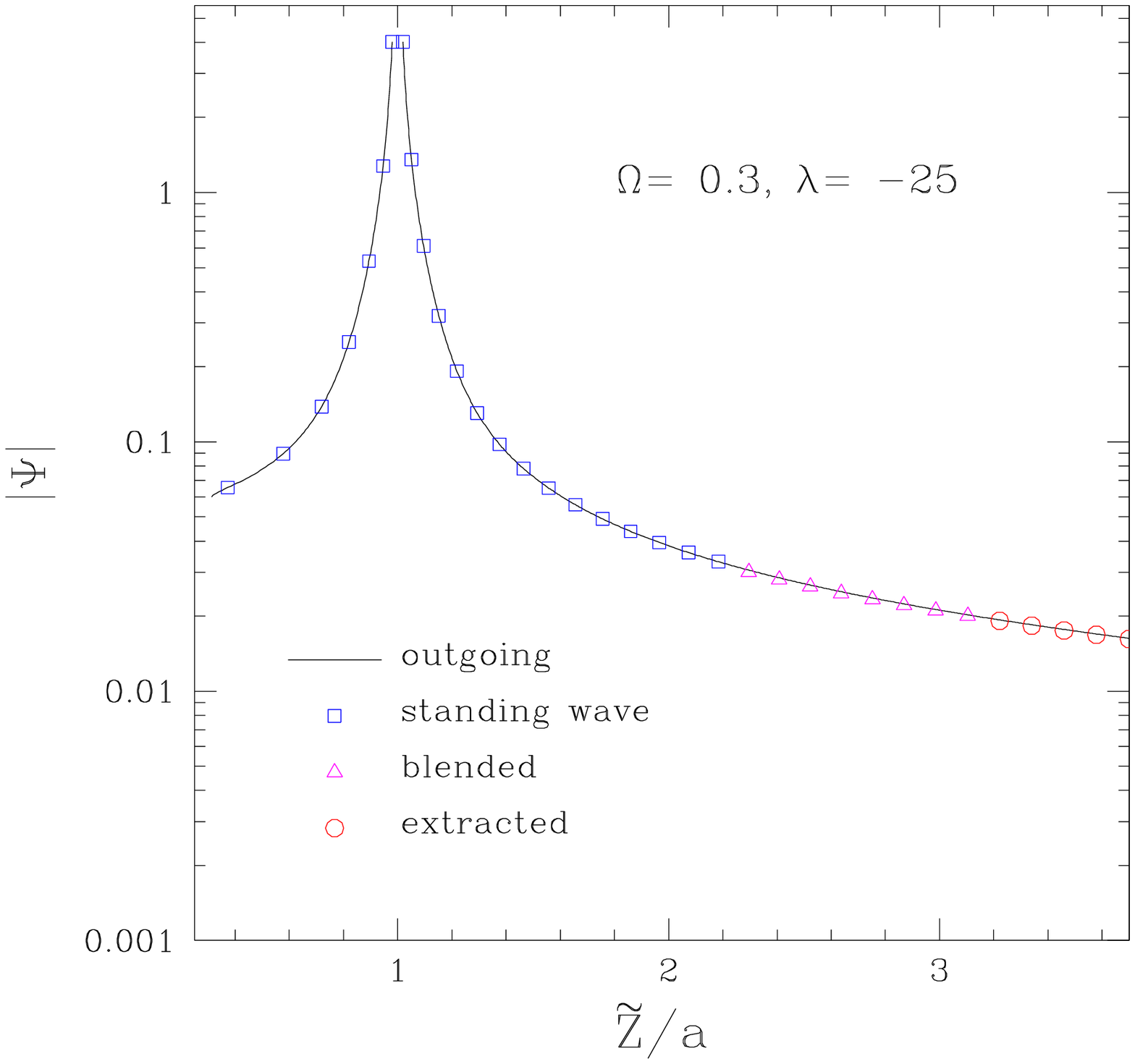}
\includegraphics[width=.4\textwidth]{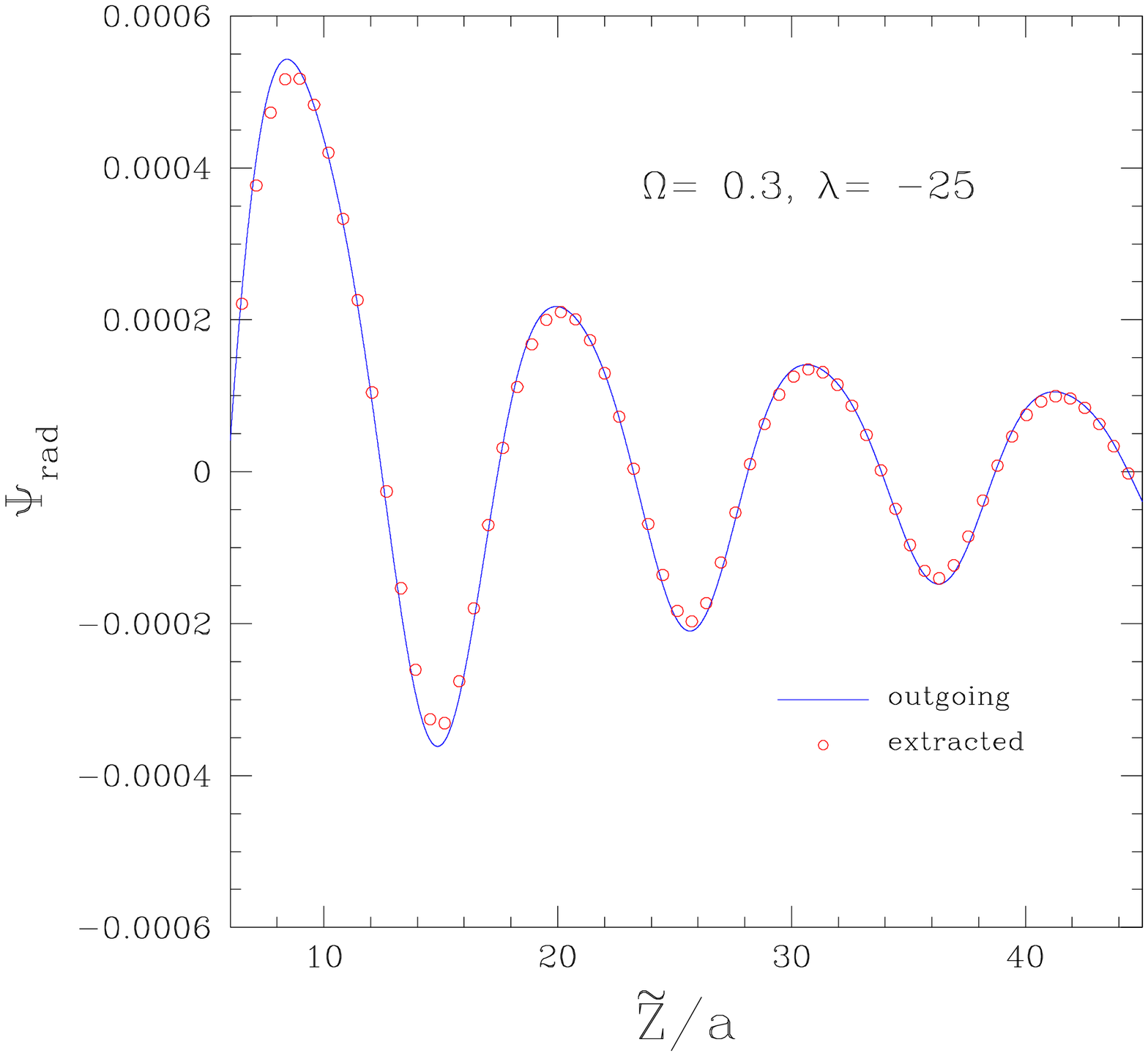} \caption{
Comparison of a computed outgoing nonlinear solution (continuous 
curve), and an approximation to the outgoing solution extracted 
from the standing wave solution (data type points).
Results are shown for a
typical
nonlinear scalar model, with parameters $\Omega=0.3 $, $\lambda=-25$,
$\Psi_0=0.15$, $\chi_{\rm min}/a=0.05$, $\chi_{\rm max}/a=200$, for a
grid with $n_{\chi}\times n_{\Theta}\times
n_{\Phi}$=$12001\times16\times32$.  Results are shown along the
$\widetilde{Z} $ axis. 
The solution is plotted as
a function of  $\widetilde{Z}$, the distance, from the origin, along the axis through 
the source.
The extracted points in the wave zone are a result of treating the waves
as linear. The small-distance plot shows the blending region
and the inner region in which the standing wave solution 
is used as an approximation for the outgoing solution.
\label{fig:extract}}
\end{figure}

In Table~\ref{table:efflin}, we present a broad overview of the
validity of effective linearity for a range of nonlinear strengths in
$a\Omega=0.3$ models.  As in table~\ref{table:yukawa}, we give the
nonlinear ``reduction factor,'' the reduction in wave amplitude due to
nonlinear effects.  Here we present a comparison of those factors for
computed outgoing solutions and for the approximate outgoing solution
extracted from the standing wave solution.  
It is clear that judged by the criterion of reduction factor
(and limited to $a\Omega=0.3$ models), effective linearity is highly
accurate, within a percent or so, for models with extremely strong
nonlinear effects. In the $\Psi_0=0.01$, $\lambda=-100$ model, the
nonlinearity reduces the wave amplitude by a factor of 40, but
effective linearity appears to be accurate to better than 1\%. 
It should be noted that the agreement of the computed outgoing
solution and the extracted outgoing solution is excellent
even for models (e.g.\,, $\Psi_0=0.01$ and small $\lambda$)
for which the strong nonlinear effects are not confined to a small
region around the source points. This is evidence that effective 
linearity does not require such confinement; it only requires that 
the nonlinearity  falls off before outer boundary effects are important, i.e.\,,
in the induction and wave zones.

\begin{table}
\caption{
The reduction factors for nonlinear outgoing waves
(the decrease in amplitude due to nonlinear effects).
For $a\Omega=0.3$ models,
the factors are compared for the directly computed outgoing
solutions and for outgoing solutions extracted from
nonlinear standing waves solutions. The $\lambda=0$
results indicate linear models in which the reduction
factor is unity by definition. The value 1.0064
found for the extracted solution gives an indication 
of the numerical accuracy of the extraction procedure.
All results were computed with $\ell_{\rm max}=3$,
$\chi_{\rm min}=0.2a$, and $\chi_{\rm max}=50a$
on a grid with $n_\chi\times n_\Theta\times n_\Phi
=8001\times16\times32$. The reduction
factor was computed by taking the ratio of the 
quadrupole components. 
Also listed are the estimated values of $R_{\rm lin}$,
the distance from the sources beyond which the nonlinear 
effects are suppressed, and  estimates of 
the reduction factors based on the estimates of 
$R_{\rm lin}$. [See Eq.~(\ref{estimate1})]
For consistency with table~\ref{table:yukawa}
the source strength has been taken to be 1.048.
\label{table:efflin} }
\begin{tabular}{|c|c||c|c|c|c||c|c|c|}
\hline
\multicolumn{1}{|c}{}&
\multicolumn{4}{c|}{$\Psi_0=0.15$}&
\multicolumn{4}{c|}{$\Psi_0=0.01$}\\
\hline 
$\lambda$&$R_{\rm lin}/a$&estimate
&true&extract&$R_{\rm lin}/a$&estimate&true&extract\\
\hline
    0  &      - &-       &    1    &  1.0064  &     - &      - & 1        &   1.0064\\
   -1  &  0.355 & 0.7012 &   0.7695 & 0.7740  &1.574  &0.2072  & 0.1745   &   0.1728\\
   -2  & 0.321 &0.6348  &   0.6813  & 0.6852  &1.266  &0.1668  & 0.1323   &   0.1315\\
   -5  & 0.274 &0.5417  &   0.5565  & 0.5597  &0.936  & 0.1233 & 0.09121  &   0.09109\\
  -10  & 0.238 &0.4707  &   0.4643  & 0.4669  &0.737  &0.09711  & 0.06764  &   0.06771\\
  -25  & 0.193 &0.3811  &   0.3548  & 0.3569  &0.532  &0.0700  & 0.04453  &   0.04466\\
  -50  & 0.162 &0.3191  &   0.2849  & 0.2865  &0.412  &0.05427  & 0.03231  &   0.03244\\
 -100  & 0.133 &0.2635  &   0.2265  & 0.2278  &0.317  &0.0418  & 0.02368  &   0.02380\\
\hline
\end{tabular}
\end{table}

\section{Conclusions}\label{sec:conc}

The fundamental concepts of the PSW method were introduced in Paper I.
In the current paper we concentrate on efficient numerical methods 
for solving the mixed PDEs of the PSW method. The innovative method
we present here is  a  mixture of adapted coordinates, multipole filtering, 
and the use of eigenvectors in place of continuum multipoles. This method seems
to meet the needs of the problem remarkably well. The method requires 
relatively little machine memory, and runs very quickly on workstations. 
The power of the method has allowed us to run nonlinear scalar field
models with larger velocity and much larger nonlinearity than was possible
with the method of Paper I.

We have shown that the method is convergent and reliable in a number
of senses: (i)~For a linear problem, the computed solution converges
to the known analytic solution as the computational grid becomes
finer and the number of retained multipoles increases. (ii)~For a
nonlinear model the Newton-Raphson iteration stably and reliably gives
a solution to outgoing or standing wave problem. We have confirmed
that our solutions agree, to the expected accuracy, with the results
presented in Paper I.

In addition to the role they play in the efficient computation, the
adapted coordinates
are very well suited to the specifications of inner boundary conditions, 
rather than to the specification of actual source terms. We have 
confirmed that there is low sensitivity to the details of the inner boundary
conditions. The solution in the wave zone has a sensitivity to these
conditions that is compatible with physical intuition; there is 
no excess sensitivity that is an artifact of the numerical method.

Two major points are worth emphasizing. First, we have confirmed that
excellent results can be obtained for moderate source velocities with
computations that keep only the monopole and the quadrupole moments of
the adapted coordinates. This allows an enormous decrease in the
computational intensity of a solution. The cost is only a moderate
increase in analytic complexity.  
A second and even more important point concerns``effective linearity,'' 
 the
approximate validity of superposing half-ingoing and half-outgoing
nonlinear solutions. We have been able to verify effective linearity
for a wider range of nonlinear models than in Paper I, including 
models with extremely strong nonlinearity.

\section{Acknowledgment} 
We gratefully acknowledge the support of NSF grant PHY0244605 and NASA
grant ATP03-0001-0027, to UTB and of NSF grant PHY-0099568 and NASA
grant NAG5-12834 to Caltech.
We thank Chris Beetle and Lior Burko for useful discussions
and suggestions. We also thank Alexey Blinov and Maria Cranor
for help with figures, and with checking calculations.


\appendix

\section{Coefficients for adapted
coordinates}\label{app:3Dcoeffs}

The inner products 
$\vec{\nabla}\chi\cdot\vec{\nabla}\Theta$,
$\vec{\nabla}\chi\cdot\vec{\nabla}\Phi$, and
$\vec{\nabla}\Theta\cdot\vec{\nabla}\Phi$, vanish since 
the adapted coordinates are orthogonal (with respect to a 
Cartesian metric on $\widetilde{X}$,$\widetilde{Y}$,$\widetilde{Z}$).
The other inner products and Laplacians are evaluated with the
explicit 
transformations in Eqs.~(\ref{chiofXYZ})--(\ref{yof}),
from which we find
\begin{eqnarray}
\nabla^2\chi&=&\frac{a^2+2Q}{\chi^3}\label{laplacechi}  \\
\nabla^2\Theta&=&{\frac {\sqrt {Q+{a}^{2}
+{\chi}^{2}\cos(2\,\Theta)}\;}{\sqrt {Q-{a}^{2}-{\chi}^{2}\cos(2\,\Theta)}}}
\;\frac{\left (Q-{a}^2\right )}{\chi^4}\\
\nabla^2\Phi&=&0
\end{eqnarray}
\begin{eqnarray}
\vec\nabla\chi\cdot\vec\nabla\chi&=&\frac{Q}{\chi^2}\\
\vec\nabla\Theta\cdot\vec\nabla\Theta&=&\frac{Q}{\chi^4}\\
\vec\nabla\Phi\cdot\vec\nabla\Phi&=&
2\;{\frac {Q+{a}^{2}+{\chi}^{2}\cos(2\,\Theta)}{{\chi}^{4}
 \sin^2(2\,\Theta)}}\label{delPhisq} 
\end{eqnarray}
where $Q$ is the function
\begin{equation}\label{Qdef} 
Q\equiv\sqrt {{a}^{4}+2\,{a}^{2}{\chi}^{2}\cos(2\Theta)+{\chi}^{4}}\ .
\end{equation}

In general, the 
$\bar{A}$ and $\bar{B}$ terms are computed from the following:
\begin{eqnarray}
\bar{A}_{\chi\chi}&=&Z^2\left(\frac{\partial\chi}{\partial X}\right)^2
+X^2\left(\frac{\partial\chi}{\partial Z}\right)^2
-2XZ\left(\frac{\partial\chi}{\partial X}\right)
\left(\frac{\partial\chi}{\partial Z}\right)\\
\bar{A}_{\Theta\Theta}&=&Z^2\left(\frac{\partial\Theta}{\partial X}\right)^2
+X^2\left(\frac{\partial\Theta}{\partial Z}\right)^2
-2XZ\left(\frac{\partial\Theta}{\partial X}\right)
\left(\frac{\partial\Theta}{\partial Z}\right)\\
\bar{A}_{\Phi\Phi}&=&Z^2\left(\frac{\partial\Phi}{\partial X}\right)^2
+X^2\left(\frac{\partial\Phi}{\partial Z}\right)^2
-2XZ\left(\frac{\partial\Phi}{\partial X}\right)
\left(\frac{\partial\Phi}{\partial Z}\right)\\
\bar{A}_{\chi\Theta}&=&Z^2\left(\frac{\partial\chi}{\partial X}\right)
\left(\frac{\partial\Theta}{\partial X}\right)
+X^2\left(\frac{\partial\chi}{\partial Z}\right)
\left(\frac{\partial\Theta}{\partial Z}\right)
-XZ\left[
\left(\frac{\partial\chi}{\partial Z}\right)
\left(\frac{\partial\Theta}{\partial X}\right)
+\left(\frac{\partial\chi}{\partial X}\right)
\left(\frac{\partial\Theta}{\partial Z}\right)
\right]\\
\bar{A}_{\chi\Phi}&=&Z^2\left(\frac{\partial\chi}{\partial X}\right)
\left(\frac{\partial\Phi}{\partial X}\right)
+X^2\left(\frac{\partial\chi}{\partial Z}\right)
\left(\frac{\partial\Phi}{\partial Z}\right)
-XZ\left[
\left(\frac{\partial\chi}{\partial Z}\right)
\left(\frac{\partial\Phi}{\partial X}\right)
+\left(\frac{\partial\chi}{\partial X}\right)
\left(\frac{\partial\Phi}{\partial Z}\right)
\right]\\
\bar{A}_{\Theta\Phi}&=&Z^2\left(\frac{\partial\Theta}{\partial X}\right)
\left(\frac{\partial\Phi}{\partial X}\right)
+X^2\left(\frac{\partial\Theta}{\partial Z}\right)
\left(\frac{\partial\Phi}{\partial Z}\right)
-XZ\left[
\left(\frac{\partial\Theta}{\partial Z}\right)
\left(\frac{\partial\Phi}{\partial X}\right)
+\left(\frac{\partial\Theta}{\partial X}\right)
\left(\frac{\partial\Phi}{\partial Z}\right)
\right]\\
\bar{B}_\chi&=&Z^2\left(\frac{\partial^2\chi}{\partial X^2}\right)
+X^2\left(\frac{\partial^2\chi}{\partial Z^2}\right)
-2XZ\left(\frac{\partial^2\chi}{\partial X\partial Z}\right)
-X\left(\frac{\partial\chi}{\partial X}\right)
-Z\left(\frac{\partial\chi}{\partial Z}\right)\\
\bar{B}_\Theta&=&Z^2\left(\frac{\partial^2\Theta}{\partial X^2}\right)
+X^2\left(\frac{\partial^2\Theta}{\partial Z^2}\right)
-2XZ\left(\frac{\partial^2\Theta}{\partial X\partial Z}\right)
-X\left(\frac{\partial\Theta}{\partial X}\right)
-Z\left(\frac{\partial\Theta}{\partial Z}\right)\\
\bar{B}_\Phi&=&Z^2\left(\frac{\partial^2\Phi}{\partial X^2}\right)
+X^2\left(\frac{\partial^2\Phi}{\partial Z^2}\right)
-2XZ\left(\frac{\partial^2\Phi}{\partial X\partial Z}\right)
-X\left(\frac{\partial\Phi}{\partial X}\right)
-Z\left(\frac{\partial\Phi}{\partial Z}\right)\ .
\end{eqnarray}

In the case of the TCBC coordinates defined in Eqs.~(\ref{chiofXYZ})
-- (\ref{yof}), the explicit forms of the coefficients are
\begin{equation}\label{barAchichi} 
\bar{A}_{\chi\chi}=
\frac{a^4\sin^2(2\Theta)\;\cos^2\Phi}{\chi^2}
\end{equation}
\begin{equation}
\bar{A}_{\Theta\Theta}=
\frac{\cos^2\Phi\;\left[\chi^2+a^2\cos(2\Theta)\right]^2}{\chi^4}
\end{equation}

\begin{equation}
\bar{A}_{\Phi\Phi}=
\sin^2\Phi\;\frac{
Q+a^2+\chi^2\cos(2\Theta)
}{
Q-a^2-\chi^2\cos(2\Theta)
}
\end{equation}

\begin{equation}
\bar{A}_{\chi\Theta}=
\frac{a^2\;\left[
\chi^2+a^2\cos(2\Theta)
\right]\;\sin(2\Theta)\cos^2\Phi}{\chi^3}
\end{equation}

\begin{equation}
\bar{A}_{\chi\Phi}=
-\frac{a^2\;\left[Q+a^2+
\chi^2\cos(2\Theta)
\right]\;\sin\Phi\cos\Phi}{\chi^3}
\end{equation}

\begin{equation}
\bar{A}_{\Theta\Phi}=
-{\frac {\sin(\Phi)\cos(\Phi)\left[ {a}^{2}+{\chi}^{2}\cos(2\,\Theta)+
Q\right]\left[{\chi}^{2}+{a}^{2}\cos(2\,\Theta)\right]}{{\chi}^{4}
\sin(2\,\Theta)}}
\end{equation}

\begin{equation}
\bar{B}_\chi=\frac{a^2\left[
\cos^2(\Phi)\left\{
3a^2\cos^2(2\Theta)-Q-2a^2+\chi^2\cos(2\Theta)
\right\}+Q+a^2+\chi^2\cos(2\Theta)
\right]}{\chi^3}
\end{equation}

\begin{equation}\label{barBPhi} 
\bar{B}_\Phi=\frac{\left(3Q+a^2+\chi^{2}\cos{2\Theta}\right)
\sin(\Phi)\cos(\Phi)}
{Q-a^2-\chi^{2}\cos{2\Theta}}
\end{equation}

\begin{equation}
\bar{B}_\Theta=\frac{
\sqrt{Q+a^2+\chi^2\cos(2\Theta)}
}{\chi^6\sqrt{Q-a^2-\chi^2\cos(2\Theta)}
}
\; (c\cos^2\Phi+d)
\end{equation}
where
\begin{equation}\label{ceq} 
c\equiv {a}^{2}{\chi}^{4}\cos(2\,\Theta)+2\,{a}^{4}{\chi}^{2}+4\,{a}^{6}\cos(2
\,\Theta)+4\,{a}^{4}{\chi}^{2}\left (\cos(2\,\Theta)\right )^{2}-4\,{a
}^{4}Q\cos(2\,\Theta)-2\,{a}^{2}Q{\chi}^{2}-{\chi}^{6}
\end{equation}
\begin{equation}\label{deq} 
d\equiv{\chi}^{4}\left ({a}^{2}\cos(2\,\Theta)+{\chi}^{2}\right )\,.
\end{equation}

The coefficients needed in the Sommerfeld boundary condition
Eq.~(\ref{FDMoutbc}) are 
\begin{eqnarray}
\Gamma^\Theta&=&\frac{\chi^2+a^2\cos{2\Theta}}{\chi^2}\;\cos{\Phi}
=\cos{\Phi}\left(1+{\cal O}(a^2/\chi^2)\right)\\
\Gamma^\Phi&=&-\sqrt{\frac{Q+a^2+\chi^2\cos{(2\Theta)}
}{Q-a^2-\chi^2\cos{(2\Theta)} \;}}\;\sin{\Phi}
=-\cot{\Theta}\sin{\Phi}\left(1+{\cal O}(a^2/\chi^2)\right)\\
\Gamma^\chi&=&\frac{1}{\chi^3}\;\sqrt{\left[Q+a^2+\chi^2\cos{(2\Theta)}
\right]\left[Q-a^2-\chi^2\cos{(2\Theta)}\right]}
=\frac{2\sin{2\Theta}}{\chi}
\left(1+{\cal O}(a^2/\chi^2)\right)\,.
\end{eqnarray}

\section{The standard spectral method for the 2+1 dimensional linear
scalar field}\label{app:SSM2D}

Here we consider the 2+1 dimensional version of our helical problem,
one equivalent to the 3+1 problem with line sources that are
infinitely long in the $\widetilde{z}$ (equivalently $\widetilde{Y}$ )
direction.  We choose to set $\lambda=0 $, i.e.\,, to make the problem
linear, since that will turn out to allow a very efficient method of
multipole projection.  The 2+1 dimensional version of
Eqs.~(\ref{origcoords}) and (\ref{ptsource})
is  
\begin{equation}\label{2+1form} 
\frac{1}{r}\frac{\partial}{\partial r}
\left(r\;\frac{\partial\Psi}{\partial r}\right)
+\left(\frac{1}{r^2}-\Omega^2\right)\,\frac{\partial^2\Psi
}{\partial\varphi^2}
=\gamma^{-1}\frac{\delta(r-a)}{a}\left[
\delta(\varphi)+\delta(\varphi-\pi)
\right]\,.
\end{equation}
Here $\varphi\equiv\phi-\Omega t$ where $\phi$ is the usual
polar angle $\tan^{-1}(\widetilde{y}/\widetilde{x})$
in the $\widetilde{x},\widetilde{y}$
plane.

The 2+1 dimensional forms of the adapted coordinates of
Sec.~\ref{sec:adapcoord} are
\begin{eqnarray}
\chi&\equiv&\sqrt{r_1r_2}
=\left\{\left[\left(\widetilde{x}-a\right)^2+\widetilde{y}^2\right]
\left[\left(\widetilde{x}+a\right)^2+\widetilde{y}^2
\right]\right\}^{1/4}\label{chiofXYZ2}\\
\Theta&\equiv&\frac{1}{2}\left(\theta_1+\theta_2\right)=
\frac{1}{2}\tan^{-1}\left(\frac{
2\widetilde{x}\widetilde{y}
}{\widetilde{x}^2-a^2-
\widetilde{y}^2
}\right)\label{ThetofXYZ2}\,.
\end{eqnarray}
With these coordinates the 2+1 dimensional version of
Eq.~(\ref{fieldtheory}) takes a form like that of
Eq.~(\ref{waveq}). As in the 3-dimensional case we use only the
homogeneous form of Eq.~(\ref{2+1form}), since the effect of the
source will be introduced through inner boundary conditions. 

In working with the sourceless linear 2+1 dimensional problem,
it turns out to be convenient to 
divide 
the original wave equation by $Q/\chi^{2}$, where
\begin{equation}
Q\equiv\sqrt {{a}^{4}+2\,{a}^{2}{\chi}^{2}\cos(2\Theta)+{\chi}^{4}\;}\ .
\end{equation}
The result is  
\begin{equation}\label{psieq} 
{\cal L}\Psi\equiv
\chi^2 Q^{-1}\,\left(\nabla^2-\Omega^2\partial^2_\varphi
\right)\Psi
={A}\;\frac{\partial^2\Psi}{\partial\chi^2}
+{B}\;\frac{\partial^2\Psi}{\partial\Theta^2}
+2{C}\;\frac{\partial^2\Psi}{\partial\chi\partial\Theta}
+{D}\;\frac{\partial\Psi}{\partial\chi}
+{E}\;\frac{\partial\Psi}{\partial\Theta}\,,
\end{equation}
where
\begin{eqnarray}
{A}&=&1-\Omega^2\;\frac{a^4\sin^2(2\Theta)}{Q}\label{wideAdef} \\
{B}&=&\frac{1}{\chi^2}\left[1-\Omega^2\;
\frac{\left(a^2\cos(2\Theta)+\chi^2\right)^2}{Q}\right]\\
{C}&=&-\Omega^2\;
\frac{a^2\sin(2\Theta)\left(a^2\cos(2\Theta)+\chi^2\right)}{Q\,\chi}\\
{D}&=&\frac{1}{\chi}\left[1-\Omega^2\;
\frac{{a}^{2}\left
(-{a}^{2}+3\,{a}^{2}\cos^2(2\,\Theta)
+2\,{\chi}^{2}\cos{(2\,\Theta)}\right)
}{Q}\right]\\
{E}&=&+\Omega^2\;
\frac{2{a}^2
\left (2\,{a}^{2}\cos(2\,\Theta)+{\chi}^{2}\right )\sin(2\,\Theta)
}{Q\,\chi^2}\,.\label{wideEdef} 
\end{eqnarray}
By dividing through by $Q$ we have put the wave equation 
in a form in which the coefficients are $\Theta$-independent
in the $\Omega\rightarrow0$ limit. With the standard method, used
below, for
projecting out multipole components of the wave equation, this 
property of the coefficients means that the  mixing of the multipoles
can be directly ascribed to the rotation.

We now expand the standing wave solution
$\Psi(\chi,\Theta)$ as
\begin{equation}\label{modesum} 
\Psi(\chi,\Theta)=\sum_{n=0,2,4\ldots{}}^N a_n(\chi)\cos{n\Theta}\,,
\end{equation}
and our equation becomes
\begin{displaymath}
\sum_{n=0,2,4\ldots{}}^N
\left[{A}\;\frac{d^2 a_n(\chi)}{d\chi^2}
-n^2{B}\;a_n(\chi)
+{D}\;\frac{d a_n(\chi)}{d\chi}\;
\right]\cos{n\Theta}
\end{displaymath}
\begin{equation}
-\,\left[2n{C}\;\frac{d a_n(\chi)}{d\chi}\;
+n{E}a_n(\chi)
\right]\sin{n\Theta}\,.
\end{equation}
Projecting with $\int_0^{2\pi}\sin{m\Theta}\cdots d\Theta$, gives
zero by symmetry; projecting with $\int_0^{2\pi}\cos{m\Theta}\cdots
d\Theta$, gives
 \begin{equation}\label{cosproj} 
\sum_{n=0,2,4\ldots{}}^N \alpha_{mn}\;\frac{d^2 a_n(\chi)}{d\chi^2}
+\beta_{mn}a_n(\chi)
+\gamma_{mn}\;\frac{d a_n(\chi)}{d\chi}=0\quad\quad m=0,2,4\ldots{}.
\end{equation}
where
\begin{eqnarray}
\alpha_{mn}&\equiv&\frac{4}{\pi}\int_0^{\pi/2}
\cos{m\Theta}\cos{n\Theta}\left[{A}(\chi,\Theta)\right]
\,d\Theta\label{alph}  \\
\beta_{mn}&\equiv&-n^2\;\frac{4}{\pi}\int_0^{\pi/2}
\cos{m\Theta}\cos{n\Theta}\left[{B}(\chi,\Theta)\right]
\,d\Theta\nonumber\\
&&-n\frac{4}{\pi}\int_0^{\pi/2}\cos{m\Theta}\sin{n\Theta}
\left[{E}(\chi,\Theta)\right]\,d\Theta
\label{bet}  \\
\gamma_{mn}&\equiv&\frac{4}{\pi}\int_0^{\pi/2}
\cos{m\Theta}\cos{n\Theta}\left[{D}(\chi,\Theta)\right]\,d\Theta\nonumber\\
&&-2n\frac{4}{\pi}\int_0^{\pi/2}\cos{m\Theta}\sin{n\Theta}
\left[{C}(\chi,\Theta)\right]\,d\Theta\,.\label{gam} 
\end{eqnarray}

When the explicit expressions for ${A}$--${E}$,
in Eqs.~(\ref{wideAdef})--(\ref{wideEdef}),
are used in Eqs.~(\ref{alph})--(\ref{gam}), the results are
\begin{equation}\label{alphmn} 
\alpha_{mn}=\epsilon_{mn}-\Omega^2\;a^4\;\frac{4}{\pi}\int_0^{\pi/2}
\cos{m\Theta}\cos{n\Theta}\;\frac{\sin^2(2\Theta)}{Q}\;d\Theta
\end{equation}
\begin{displaymath}
\beta_{mn}=-\frac{n^2}{\chi^2}\epsilon_{mn}
+\frac{4}{\pi}\frac{\Omega^2}{\chi^2}\left[n^2\int_0^{\pi/2}
\cos{m\Theta}\cos{n\Theta}\;
\frac{\left(a^2\cos(2\Theta)+\chi^2\right)^2}{Q}\;d\Theta
\right.
\end{displaymath}
\begin{equation}\label{betmn} 
\left. -n\int_0^{\pi/2}
\cos{m\Theta}\sin{n\Theta}\;
\frac{2{a}^2
\left (2\,{a}^{2}\cos(2\,\Theta)+{\chi}^{2}\right )\sin(2\,\Theta)
}{Q}
\right]
\end{equation}
\begin{displaymath}
\gamma_{mn}=\frac{1}{\chi}\epsilon_{mn}
+\frac{4}{\pi}\frac{a^2\Omega^2}{\chi}\left[
\;-\int_0^{\pi/2}\cos{m\Theta}\cos{n\Theta}
\frac{\left
(-{a}^{2}+3\,{a}^{2}\cos^2(2\,\Theta)
+2\,{\chi}^{2}\cos{(2\,\Theta)}\right)
}{Q}\,d\Theta
\right.
\end{displaymath}
\begin{equation}\label{gammn} 
\left.
+2n\int_0^{\pi/2}\cos{m\Theta}\sin{n\Theta}
\frac{\sin(2\Theta)\;\left(a^2\cos(2\Theta)+\chi^2\right)}{Q}
\;d\Theta\right]
\end{equation}
\bigskip\bigskip
where
\begin{equation}
\epsilon_{mn}\equiv\delta_{m+n,0}+\delta_{m,n}
=\left\{
       \begin{array}{l}
       2\mbox{  if $m=n=0$} \\
       1 \mbox{  if $m=n\neq0$}\\ 
       0 \mbox{  if $m\neq n$}
              \end{array}
\right.\ \ .
\end{equation}

The integrals needed in Eqs.~(\ref{alphmn})--(\ref{gammn}) are all of the form 
\begin{equation}\label{Iintegrals} 
{\cal I}=\int_0^{\pi/2}
\frac{\cos{(2P\Theta)}\sin{(2J\Theta)}}{Q(\chi,\Theta)}\;d\Theta
\end{equation}
where $P$ and $J$ are integers. With trigonometric identities, and with the
substitution $x=\sin{\Theta}$, such integrals can all be expressed 
in terms of integrals of the form
\begin{equation}
{\cal K}_N(k)\equiv\int_0^1\frac{x^N}{\sqrt{1-k^2x^2}\sqrt{1-x^2}}\;dx\,,
\end{equation}
where
\begin{equation}
k\equiv\frac{2a\chi}{a^2+\chi^2}\ .
\end{equation}
For the integrals in Eq.~(\ref{Iintegrals}) only even values of $N
$ are needed for ${\cal K}_N$. 
All such integrals can be evaluated in terms of the complete 
elliptic integrals\cite{abramandstegun},
\begin{equation}
K(k)\equiv\int_0^1\frac{1}{\sqrt{1-k^2x^2}\sqrt{1-x^2}}\;dx\quad\quad
E(k)\equiv\int_0^1\frac{\sqrt{1-k^2x^2}}{\sqrt{1-x^2}}\;dx\ .
\end{equation}

To use the elliptic integrals to evaluate the ${\cal K}_N(k)$, for
even $N$, we start with the relationship (for $M\geq0$)
\begin{displaymath}
0=\int_0^1
(d/dx)\left[x^{2M+1}
\sqrt{1-k^2x^2}\sqrt{1-x^2}\right]
\;dx
\end{displaymath}
\begin{equation}
=\int_0^1
\frac{x^{2M}\left[
(2M+1)-(2M+2)(1+k^2)x^2+(2M+3)k^2x^4
\right]}
{
\sqrt{1-k^2x^2}\sqrt{1-x^2}}
\;dx\ .
\end{equation}
This gives us the recursion relation
\begin{equation}
(2M+1){\cal K}_{2M}(k)-
(2M+2)(1+k^2){\cal K}_{2M+2}(k)+
(2M+3)k^2{\cal K}_{2M+4}(k)=0\ .
\end{equation}
We know that 
\begin{displaymath}
{\cal K}_{0}(k)=K(k)\ ,
\end{displaymath}
and we can easily show that
\begin{displaymath}
{\cal K}_{2}(k)=\frac{1}{k^2}\left[
K(k)-E(k)
\right]\ ,
\end{displaymath}
so all values of ${\cal K}_{2M}(k)$ follow from the 
recursion relation.

Very efficient computation follows from the results above. 
At a given value of $\chi$, only two integral evaluations
must be done, those for $K(k)$
and $E(k)$. All values of ${\cal K}_{2M}(k)$ then follow from the recursion
relation, and hence all values of $\alpha_{mn}$, $\beta_{mn}$, $\gamma_{mn}$,
can be found at negligible computational expense.

Here we use this method only to illustrate important general
issues. To make this illustration as clear as possible, 
we take the simplest nontrivial case of the multipole
expansion: we keep only the $n=0$  and $n=2$
terms\cite{alphbetgam}, so that 
\begin{equation}
\Psi(\chi,\Theta)= a_0(\chi)+
a_2(\chi)\cos{2\Theta}\ .
\end{equation}
The equations in Eq.~(\ref{cosproj}), then describe the interaction of
the monopole $a_0(\chi)$ and quadrupole $a_2(\chi)$ terms.  For
$\chi/a\ll1$ and $\chi/a\gg1$ the terms that mix the multipoles die
off; it is only in the transaction region, $\chi/a\sim1$ that there is
strong mixing of the multipoles, a mixing that for our problem is
quadratic in the source velocity $a\Omega$. The process of the
generation of radiation can be viewed as the growth of $a_2$ from the
small-$\chi$ near-source region to the large-$\chi$ radiation region.

In principle, the linear standing wave problem could be solved by
including a term $a_s\sin(2\Theta)$, leading to an additional second
order differential equation.
At some inner boundary $\chi_{\rm min}\ll a$ the values could be
specified for all multipoles, and at some outer boundary $\chi_{\rm
min}\gg a$, a fall off condition could be specified for $a_0$. Ingoing or
outgoing conditions could be used to relate $a_2$ and $a_s$.

It is easier, and more instructive, to use another approach to finding
the standing wave solution, the minimum amplitude 
method presented in Paper I. For the linear problem it is
straightforward to show\cite{WBLandP} that of all solutions that (i)
have the form of Eq.~(\ref{modesum}), and (ii) correctly couple to the
source, the solution with the minimum wave amplitude in each multipole
is the standing wave solution, i.e.\,, the solution that is
half-ingoing and half-outgoing. 
In principle, the minimum-amplitude criterion can be used as a definition
of standing waves in a nonlinear problem. Here we are dealing with
a linear problem, so we are simply exploiting a known property 
of the solutions.

In this minimum-amplitude method we specify $a_0$, $da_0/d\chi$, $a_2$
and $da_2/d\chi$ at $\chi_{\rm min}$, then shoot outward. The choice of
$a_0$ and $da_0/d\chi$, at $\chi_{\rm min}$ are those for unit point
charges.  The value of $a_0$ at $\chi_{\rm min}$ sets the scale of the
linear solution. The value of $da_0/d\chi$ can be approximated as
$1/\pi\chi_{\rm min}$. In principle this starting value can be
adjusted so that at large $\chi$ the results for $a_0$ satisfy the
fall-off condition $da_0/d\chi=a_0/\chi\log{\chi} $. In practice, the
overall result (after the minimization described below) is very
insensitive to the choice of $da_0/d\chi$ at $\chi_{\rm min}$.

The choices 
for $a_0$ and $da_0/d\chi$ at $\chi_{\rm min}$
are determined by the form of $\Psi$
very close to one of the unit point charges. 
Some care is necessary in taking this limit.
In the $x,y$ frame comoving with one of the unit point charges
the value of $\Psi$
 due only to that 
charge is $\log{[(x^2+y^2)/a^2]}/4\pi$. 
We must now transform this result to the ``lab''
frame of our calculation in which we use coordinates
$\widetilde{x},\widetilde{y},t$. 
The  Lorentz 
transformation is
\begin{equation}
x=\widetilde{x}\quad\quad\quad y=\gamma\widetilde{y}\,.
\end{equation}
(The last relation follows since $\widetilde{y} $ is in a coordinate
frame that comoves with the point source, but which is related to the
lab frame by a simple translation by $vt $.)
Since $\Psi
$ is a Lorentz invariant we have 
$\Psi=\log{(r^2
/a^2)}/4\pi$ where $r^2=\widetilde{x}^2+\gamma^2\widetilde{y}^2
$, expressed in adopted coordinates, is
\begin{equation}
r^2=\frac{1}{4}\,\frac{\chi^4}{a^2}\left[
1-\frac{1}{2}\frac{\chi^2}{a^2}\cos{(2\Theta)}
\right]+\left(\gamma^2-1
\right)\,\frac{1}{4}\sin^2{(2\Theta)}
\frac{\chi^4}{a^2}\left[1-\frac{\chi^2
}{a^2
}\cos{2\Theta}\right]\,,
\end{equation}
and $\Psi$ is therefore
\begin{displaymath}
\psi=\frac{1}{4\pi}\left(
\log{\left[\frac{\chi^4}{4a^4}\right]}
+\log{
\left[1+\left(\gamma^2-1\right)
\sin^2{(2\Theta)}
\right]}
\right)+{\cal O}(\chi/a)^2
\end{displaymath}
\begin{equation}\label{fullr2} 
=\frac{1}{4\pi}\left(
\log{\left[\frac{\chi^4}{4a^4}\right]}
+2\log{
\left[\frac{\gamma+1}{2}
\right]}
\right)
+{\cal O}(\cos{(4\Theta)})
+{\cal O}(\chi/a)^2\,.
\end{equation}
The last relationship is meant to emphasize that in the
$\chi\rightarrow0$ limit $\Psi$ has no $\cos{(2\Theta)}$
component. This comes from the fact that the Lorentzian
pancaking of the source field has the nature of a local
quadrupole deformation, while the $\cos{(2\Theta)}$ term
represents a local dipole.
(As defined in
Eq.~(\ref{ThetofXYZ2}), $\Theta\approx\theta/2$ or $(\theta+\pi)/2$
near the source points, so the $\cos{(2\Theta)} $ dependence near the
source point corresponds to the dipole field of the source.)

The contribution, near $\chi=0$, due to the distant point can be
approximated with $r=2a $ so that at $\chi=\chi_{\rm min}$ the appropriate
starting conditions for unit point charges are
\begin{equation}
a_
0=\frac{1}{4\pi}\,\left(\log{\left[\frac{\chi^4}{4a^4}\right]}
+2\log{
\left[\frac{\gamma+1}{2}
\right]}+\log{4}\right)
\quad
\frac{da_0}{d\chi}=\frac{1}{\pi\chi}
\quad a_2=0\quad \frac{da_2}{d\chi}=0
\end{equation}
While $da_2/d\chi$ at $\chi\rightarrow0$ vanishes in principle,
in the numerical computation,  $da_2/d\chi$ plays a more delicate 
role. It is chosen to minimize the wave amplitude at large $\chi$.  It
is actually the values of $a_2$ and $da_2/d\chi$, at $\chi_{\rm min}$
that determine the radiation field at large $\chi$, and determine
whether there is any $a_2$ radiation except the radiation coupled to
the source. To minimize that radiation (and suppress radiation that is
not coupled to the source) we fix $a_2$ and vary $da_2/d\chi$. (We
could just as well fix $da_2/d\chi$ and vary $a_2$.) 
Minimization is taken to mean the minimum value of the amplitude
defined by
\begin{equation}
{\rm Amp}\equiv\left\langle(\chi/a)^{1/2}
\sqrt{
a_2^2+\frac{1}{4\Omega^2}\left(\frac{da_2}{d\chi}+\frac{a_2
}{2\chi}\right)^2\;}\ \right\rangle\,.
\end{equation}
For the expected large distance form of the waves $
r^{-1/2}\cos{(2\Omega r+\delta})$, the quantity inside the angle
brackets is expected to be nearly $\chi$-independent; this is
confirmed by the numerical results.  The angle brackets denote an
average over a wavelength, resulting in a quantity that is $\chi$
independent to high accuracy at large $\chi$. 
In our minimization procedure we vary  $da_2/d\chi$ 
at $\chi_{\rm min}$ to minimize Amp. 
The meaningfulness of this minimization procedure, of course, 
depends on its insensitivity to the details of how 
$a_2$  and $da_2/d\chi$ are chosen at 
$\chi_{\rm min}$. This is discussed below.
\begin{figure}[ht] 
\begin{center}
\includegraphics[width=.4\textwidth]{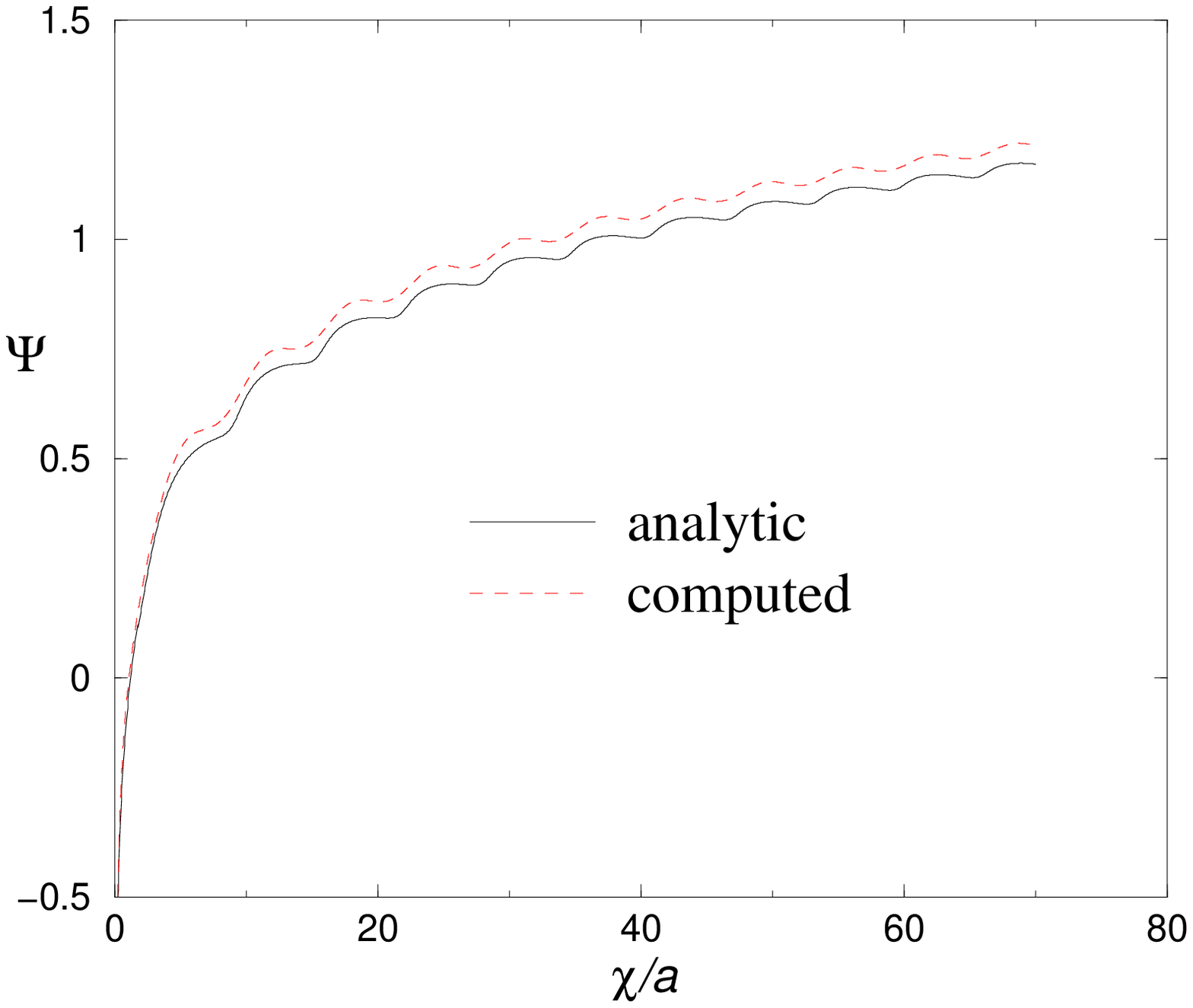} \hspace{.5in}
\includegraphics[width=.4\textwidth]{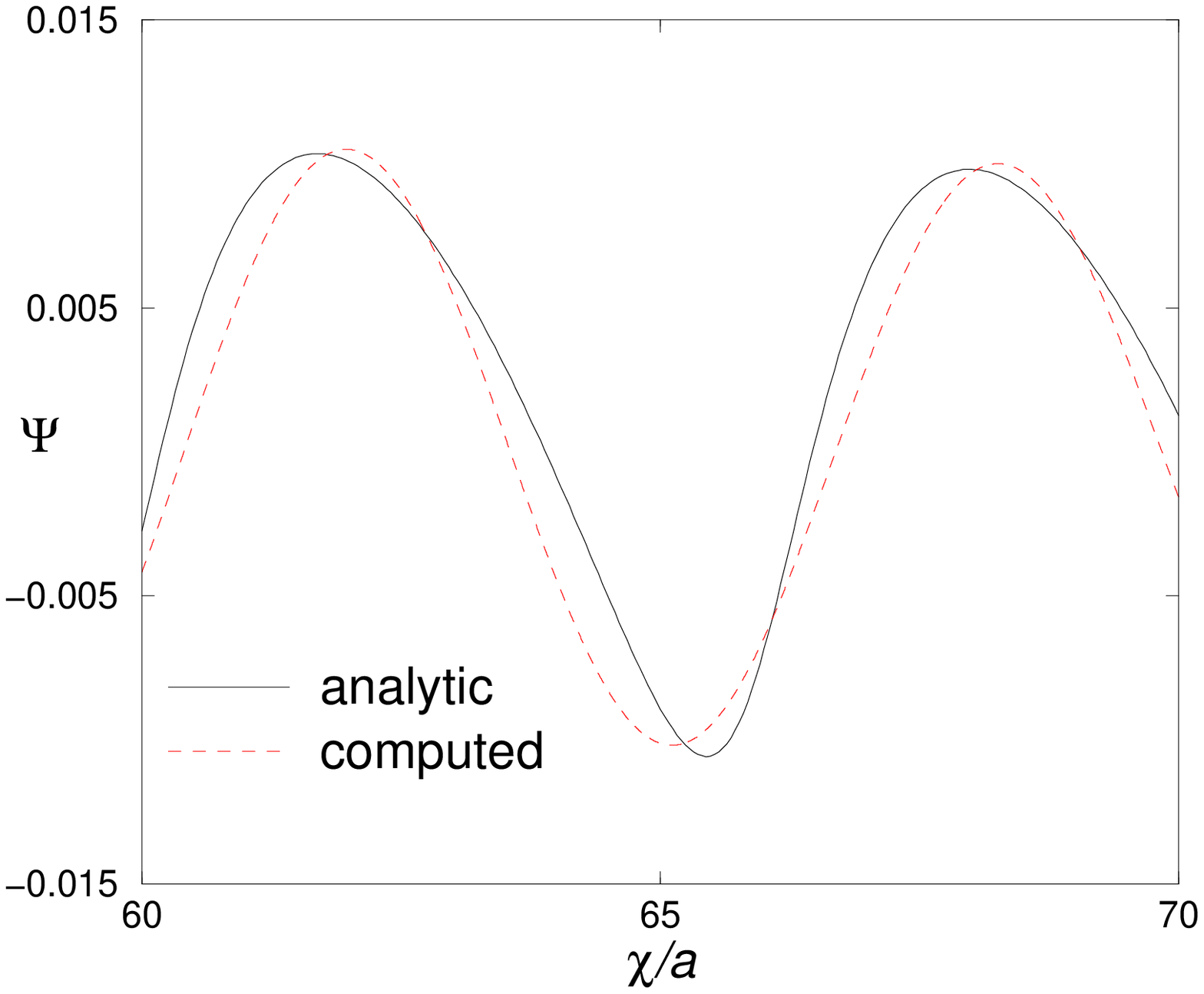} \caption{ Comparison
of the analytic waves and the computed waves for $a\Omega=0.5$,
$\chi_{\rm min}=0.05a$ and $\chi_{\rm max}=70a$, and $a_2=0
$ at $\chi_{\rm min}
=0$.  The monopole is
subtracted in the figure on the right to allow for a comparison of
computed and analytic wave amplitudes.  .  \label{fig:2Dresults}}
\end{center}
\end{figure}

For the source speed $a\Omega=0.5$, results, for $\Theta=0
$, of our computation
are shown in 
Fig.~\ref{fig:2Dresults}, and are compared with the analytic solution
for two  point sources each of unit strength 
\begin{equation}\label{besseries}
\Psi=
\sqrt{1-a^2\Omega^2\;}\left[\left\{
\begin{array}{ll}
\frac{1}{\pi}\ln{(r/a)}&\mbox{for $r>a$}\\
0&\mbox{for $r<a$}\\
\end{array}
\right\}
+\sum_{m=2,4,6\ldots{}}J_m(m\Omega r_<)
N_m(m\Omega r_>)\,\cos{(m\Theta)}\right]\,,
\end{equation}
(where $r_>,r_<$ indicate, respectively, the greater and lesser of
$r,a$).  For our more typical choice $a\Omega=0.3$, the difference of
the computed and analytic solution are too small to show up well in a
plot.  The plot on the left shows the comparison for the whole range
of the computation. Dividing the analytic solution by 1.039 brings it
into nearly perfect alignment with the computed solution; we therefore
characterize the overall error in the computed solution as 3.9\%. 
Overall errors for several values of $a\Omega$ are listed
in Table \ref{table:2Derror}.

The
plot on the right in Fig.~\ref{fig:2Dresults} focuses on the
oscillations in the far zone by removing effective monopole
terms. For the analytic result this is done by subtracting the
asymptotic monopole solution $\sqrt{1-a^2\Omega^2 \;}\log{(\chi^2/2a^2
)}/\pi$. For the computed solution this is done by plotting only
$a_2(\chi)$.  The comparison shows two effects. First, there is a 1.2\%
difference in peak-to-peak amplitude. Errors of this type are 
tabulated for different values of  $a\Omega$ in Table \ref{table:2Derror}.
A second and more interesting effect, apparent in the waves
of Fig.~\ref{fig:2Dresults}, is
the difference in shapes. Since the computed wave
contains only the quadrupole component it has a nearly perfect
sinusoidal form. The analytic solution, on the other hand, shows a
rapid rise and a slow fall off due to the contribution of the
hexadecapole and higher modes. For small $a\Omega$ the amplitude of a
$\cos{(m\Theta)}$ component depends on $a\Omega$ approximately according to
$(a\Omega)^{m-0.5}$, so the contributions from $m=4,6,8\ldots{}$ 
decrease quickly with $\Omega$. 
\begin{table}
\caption{The errors (differences from analytic solution)
of the monopole+quadrupole approximation for different values
of the source speed $a\Omega$. The overall error is the difference
of the computational and analytic solution at $\chi_{\rm max}
=70a$; the wave error is the difference of the computational
and analytic peak-to-peak amplitudes.}  \label{table:2Derror}
\begin{tabular}{|c|c|c|}
\hline
$a\Omega$& Overall Error & Wave Error\\
\hline
0.1&0.08\%&0.14\%   \\
0.3&1.0\%&0.1\%  \\
0.5&3.9\%&1.2\%  \\
0.7&13\%&3\%  \\
\hline
\end{tabular}
\end{table}

We now want to use the accuracy of the linear 2+1 dimensional model to
make an important point about the role played by the source structure
in generating radiation. To do this in our monopole+quadrupole
computational models we depart from considering ``point'' sources and
we vary the choice of $a_2(\chi_{\rm min})$. In each case the
specification of $da_2/d\chi$ at $(\chi_{\rm min})$ is fine-tuned to
get a minimum wave amplitude at $\chi_{\rm max}$. For this
investigation to have physical relevance we need to ask what a
``reasonable'' value is for $a_2(\chi_{\rm min})$.

If $\chi_{\rm surf}$ specified the (approximately spherical) location
of the outer surface of the source, then we can write
\begin{equation}
a_2(\chi_{\rm surf})=\kappa\,a_0(\chi_{\rm surf})\,,
\end{equation}
where $\kappa$ indicates the relative quadrupole strength
of the source. We expect $\kappa$ to be small for realistic
sources and of order unity only for highly distorted sources.

In our computational models we take $\chi_{\rm min}/a$ to be very small,
typically 0.05. It is useful, however, to consider the computed
results applying to larger sources, sources with $\chi_{\rm surf}$
significantly larger than $\chi_{\rm min}$. To do this we can use the
fact that outside, but very close to a source, $a_0$ varies as
$\log{\chi} $ and $a_2$ falls off as $1/\chi^2$. From this we conclude
that for small $\chi$
\begin{equation}
\frac{a_2(\chi_{\rm min})}{a_0(\chi_{\rm min})}=\kappa\;
\frac{\chi^2_{\rm surf}\log{\chi_{\rm surf}}}
{\chi^2_{\rm min}\log{\chi_{\rm min}}}\,.
\end{equation}
We can now fix $\kappa=\pm1$, its maximum reasonable value, and we can
choose a value of $\chi_{\rm surf}$, the value at which the quadrupole
and monopole have equal strength.  These choices determine
$a_2(\chi_{\rm min})/a_0(\chi_{\rm min})$, and therefore the
computational model.

Results are presented in table~\ref{table:2Dsns}. For the choices
$\chi_{\rm surf}/a =0.1$, 0.2 and 0.3, and for $\kappa=$1 and -1, the
amplitude of the quadrupole waves is computed for $\chi_{\rm
min}/a=0.05$ and $\chi_{\rm max}/a=70$. For each choice of
$a_2(\chi_{\rm min})$, the starting value of $da_2/d\chi$ is fine
tuned for minimum wave amplitude. Since $a_2(\chi)$ should have the
form const./$\chi^2$, the value of $da_2/d\chi$ should be very nearly
equal to $-2a_2/\chi$. The values for $da_2/d\chi$ in
table~\ref{table:2Dsns} are very close to this prediction.  The table
presents the ``relative amplitude,'' the ratio of the radiation
amplitude to the radiation amplitude for the case $a_2(\chi_{\rm min}
)=0$. It is clear from the results that the structure of the source
has little influence on the radiation unless the source is large and
has an extreme nonspherical deformation.

In the case of gravitational sources, an even stronger statement can
be made.  As already point out, the ``quadrupole'' mode
$a_2(\chi)\cos(2\Theta)$ near the source point is actually a local
dipole $a_2(\chi)\cos(\theta_1)$.  This dipole moment can be viewed as
a displacement of the center of scalar charge, and hence a change in
the radius at which the source points move. The change in the
radiation amplitude can be ascribed to this change in radius. This
explanation of the change can be made quantitative. We let $r$
represent the distance from source point 1 at which inner boundary
data is being specified and we let $\delta$ be the radial distance by
which the center of scalar charge is being moved outward. The solution
for the scalar field is
\begin{equation}
\Psi=\frac{1}{4\pi}\log{\left(r^2-2r\delta\cos{\theta_1}
+\delta^2\right)}
\approx\frac{1}{4\pi}\left[\log{r^2}
-2(\delta/r)\cos{\theta_1}
\right]\,.
\end{equation}
From this we infer that 
\begin{equation}
a_2=-
\frac{\delta}{2\pi r}\,.
\end{equation}
This means that the radius of orbital motion is changed
from $a$ according to
\begin{equation}
a\rightarrow a+\delta=a-2\pi ra_2\,.
\end{equation}
From Eq.~(\ref{besseries}) we see that for quadrupole 
radiation (that is, $m=2$) the amplitude of the 
waves scales $\propto a^2$. It follows that the 
amplitude of the waves should depend on $a_2$
according to 
\begin{equation}
{\rm amplitude}\propto 1-4\pi r a_2/a\,.
\end{equation}
For the computations presented in table~\ref{table:2Dsns}, $a_2$ is
evaluated at $\chi_{\rm min}=0.05a$, so $r\approx0.05^2a$.  The
numerical results following from this simple explanation of the shift
of the center of charge, presented as the last column in
table~\ref{table:2Dsns}, are convincingly accurate.

This explanation for the role of the $m=2$ mode is of considerable
significance for gravitational problems. The equivalence principle
implies that there is no local dipole for the gravitational sources.
Thus the starting value of $a_2$ which, at large distances gives the 
quadrupole radiation, is not a parameter of the structure of the 
source; if we know the location of the effective center of the source,
the structure is fixed.

\begin{table}
\caption{The effect on the wave amplitude of the conditions on $a_2$
and $da_2/d\chi$ at $\chi_{\rm min}=0.3a$. The role of the $a_2$
term can be understood as an effective shift of the center 
of scalar charge. See text for details.}  \label{table:2Dsns}
\begin{tabular}{|c|c|c|c|c|c|}
\hline
$\kappa$& 
$\chi_{\rm surf}$& 
$a_2(\chi_{\rm min})$&
$da_2/d\chi\ {\rm at}\ \chi_{\rm min}$&Rel.~Amp.&$1-4\pi ra_2/a
$
\\
\hline
1&0.1&-2.92007&116.999652&1.045&1.046\\
-1&0.1&2.92007&-117.011864&0.955&0.954\\
1&0.2&-8.16416&327.127955&1.125&1.128\\
-1&0.2&8.16416&-327.140167&0.875&0.872\\
1&0.3&-13.7416&550.612310&1.211&1.216\\
1&0.3&13.7416&-550.6245521&0.789&0.784\\
\hline
\end{tabular}
\end{table}

\section{Details of the Eigenspectral method}\label{app:ESMdetails}

In this appendix we explain how the continuum angular Laplacian of
Eq.~(\ref{angLap}) is implemented as a  linear 
operator in the $N=n_\Theta\times n_\Phi $ dimensional space.  That
linear operator must represent 
the angular Laplacian evaluated at a grid point $\Theta_a$,
$\Phi_b$. That is, the linear operator $L_{ab,ij}$ must satisfy
\begin{equation}\label{anglapdef} 
\left[\sin\Theta\nabla_{\rm ang}^2\Psi\right]_{ab}
\approx \sum_{ij}
L_{ab,ij}\Psi_{ij}\,,
\end{equation}
where the approximation is due to FDM truncation error.  It will be
convenient below to write the linear operator as a sum ${L}={
L}^{(1)}+{L}^{(2)}$ with ${ L}^{(1)}$ containing $\Theta$
derivatives and ${L}^{(2)}$ containing $\Phi$ derivatives.

By exploiting the symmetries of the PSW configuration we can
limit the range of angular coordinates to one quarter of the complete
2-sphere. 
The indices $a,i$ range from 1 to $n_\Theta$ , representing,
respectively $\Theta=\Delta\Theta/2$ to
$\Theta=\pi/2-\Delta\Theta/2$.
The indices  $b,j$ range from 1 to
$n_\Phi$, representing,
respectively $\Phi=\Delta\Phi/2$ to
$\Phi=\pi-\Delta\Phi/2$.
Our goal here will be to show that with this choice of the grid, the 
linear operator 
can be chosen to have the symmetry
\begin{equation}\label{defsym} 
L_{ab,ij}=L_{ij,ab}\,.
\end{equation}

The 
elements of ${L}_{ab,ij}
$ have a different form for the boundaries at
$\Theta=\Delta\Theta/2$, $\pi/2-\Delta\Theta/2$ and at
$\Phi=\Delta\Phi/2$, $\pi-\Delta\Phi/2$, and in the interior of the
angular grid. We consider each case separately.

\subsubsection*{Case I: Interior points, $1<a<n_\Theta$, $1<b<n_\Phi$}

The first contribution is
\begin{equation}
L^{(1)}_{ab,ij}=\frac{\delta_{bj}}{(\Delta\Theta)^2
}\left[\sin\Theta_{a-1/2}\delta_{a-1,i}
-\left(
\sin\Theta_{a+1/2}+\sin\Theta_{a-1/2}
\right)\,\delta_{ia}
+\sin\Theta_{a+1/2}\delta_{a+1,i}\right]
\end{equation}
and the second is
\begin{equation}
L^{(2)}_{ab,ij}=\frac{1}{(\Delta\Phi)^2
}\frac{\delta_{ai}}{\sin\Theta_a}\left(
\delta_{b-1,j}
-2\delta_{b,j}
+\delta_{b+1,j}
\right)
\,.
\end{equation}
Here $\delta_{j,c}
$ is the Kronecker delta, and $\sin\Theta_{a\pm1/2}$ is
defined to mean $\sin{(\Theta_a\pm\Delta\Theta/2 )}$.
It can be seen that both contributions to ${L}_{ab,ij}
$
are symmetric with respect to the interchange of the pair $ab$
with the pair $ij$, 
and hence Eq.~(\ref{defsym}) is satisfied.

\subsubsection*{Case II: Boundary at $a=1$, $1< b<n_\Phi$}

The $\Theta$ derivative part of the operator formally takes 
the form
\begin{equation}
L^{(1)}_{1b,ij}=\frac{\delta_{bj}
}{(\Delta\Theta)^2
}\left[\sin{0}\,\delta_{0,i}
-\left(
\sin{0}+\sin\Delta\Theta
\right)\,\delta_{i1}
+\sin\Delta\Theta\delta_{2,i}\right]\,.
\end{equation}
In the sum in Eq.~(\ref{anglapdef}) the $\delta_{0,i}$ term respresents
$\Psi(\Theta=-\Delta\Theta/2,\Phi)$, which is not a value available on the
angular grid. This term however, is multiplied by $\sin{0}=0$
and can be ignored, so $L^{(1)}_{ab,ij}$ satisfies the symmetry 
condition in Eq.~(\ref{defsym}).
The form of $L^{(2)}_{ab,ij}$ also
applies without change to $a=1$, and hence 
Eq.~(\ref{defsym}) is satisfied for the index range $a=1$, $1<b<n_\Phi$.
\begin{figure}[ht] 
\begin{center}
\includegraphics[width=.4\textwidth]{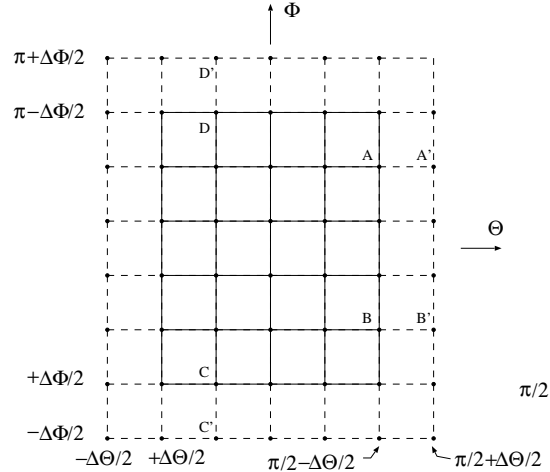} \caption{ 
An angular grid with $n_\Theta=5$
and $n_\Phi=6$. Grid points, points at which a value for $\Psi$ is computed, are connected by solid lines. The dashed lines extend the grid to ``phantom'' points
needed for the computation. For the FDM implementation of the Laplacian
at point $A$ the value of $\Psi$ at point $A'$ is needed. By the symmetry 
of the physical problem, this value can be replaced by the value 
at point $B$, which {\em is} on the grid. Similarly the value at $B'$,
when needed, can be replaced by that at point $A$; 
the value at $C'$ can be replaced by that at $C$; the value at $D'$ can be replaced by that at $D$; and so forth.
\label{fig:grid}}
\end{center}
\end{figure}
\subsubsection*{Case III: Boundary at $a=n_\Theta$, $1<b<n_\Phi$}

In this case the $\Theta$ derivative part of the operator formally takes 
the form
\begin{equation}\label{L1boundary} 
L^{(1)}_{n_\Theta
b,ij}=\frac{\delta_{bj}}{(\Delta\Theta)^2
}\left[\sin{(\pi/2-\Delta\Theta)}\delta_{n_\theta-1
,i}
-\big(
\sin{(\pi/2-\Delta\Theta)}+\sin{(\pi/2)}
\big)\,\delta_{n_\Theta,i}
+\sin{(\pi/2)}\delta_{n_\theta+1
,i}\right]\,.
\end{equation}
The 
$\delta_{n_\theta+1,i}$
represents $\Psi(\Theta_{n_\Theta+1}
,\Phi_b)=\Psi(\Theta=\pi/2+\Delta\Theta/2,\Phi)$. This value 
is not directly available on the grid, but we can get an equivalent
value that is on the grid by using the symmetry
\begin{equation}
\Psi(\Theta,\Phi)=\Psi(\pi-\Theta,\pi-\Phi)\,.
\end{equation}
This is equivalent to the statement that $\Psi
$ is invariant with respect to inversion through the origin and 
reflection in the orbital plane (or equivalently 
$\widetilde{Z}\rightarrow-\widetilde{Z}$,
$\widetilde{X}\rightarrow-\widetilde{X}$,
$\widetilde{Y}\rightarrow\widetilde{Z}$
in 
Fig.~\ref{fig:2n3Ros}.) As shown in Fig.~\ref{fig:grid},
we can, therefore, replace 
$\Psi(\Theta=\pi/2+\Delta\Theta/2,\Phi)$
with
$\Psi(\Theta=\pi/2-\Delta\Theta/2,\pi-\Phi)$, or 
can replace 
$\Psi(\Theta_{n_\Theta+1},\Phi_{b})$
with
$\Psi(\Theta_{n_\Theta},\Phi_{\bar{}b})$
where $\bar{b}\equiv N_\Phi-b$.
Equivalently, we can rewrite 
Eq~(\ref{L1boundary}) as 
\begin{equation}\label{L1boundary2} 
L^{(1)}_{n_\Theta b,ij}=\frac{\delta_{bj}}{(\Delta\Theta)^2
}\left[\sin{(\pi/2-\Delta\Theta)}\delta_{n_\theta-1
,i}
-\big(
\sin{(\pi/2-\Delta\Theta)}+\sin{(\pi/2)}
\big)\,\delta_{n_\Theta,i}\right]
+
\frac{\delta_{\bar{b}j}}{(\Delta\Theta)^2
}
\sin{(\pi/2)}\delta_{n_\theta,i}\,.
\end{equation}
The term that has been introduced is 
\begin{equation}
L^{(1)}_{n_\Theta b,n_\Theta \bar{b}}=\frac{1}{(\Delta\Theta)^2 }\,.
\end{equation}
Since $\bar{\bar b}=b$, we have 
$L^{(1)}_{n_\Theta b,n_\Theta \bar{b} }=L^{(1)}_{n_\Theta \bar{b},n_\Theta {b}}\,$
which satisfies the symmetry in Eq.~(\ref{defsym}).
All other terms in $L^{(1)}_{n_\Theta b,ij}$ remain the same as in 
Case I, and hence
Eq.~(\ref{defsym}) is satisfied for the index range $a=n_\Theta$, $1<b<n_\Phi$.

\subsubsection*{Case IV: Boundaries at $1\leq a<n_\Theta$, $b=1$ and $b=n_\Phi$}

For these boundary points Case I considerations apply to $L^{(1)}_{ab,ij}$. 
For $b=1$, however, $L^{(2)}_{ab,ij}$ takes the form  
\begin{equation}
L^{(2)}_{a1,ij}=\frac{1
}{(\Delta\Phi)^2
}\frac{\delta_{ai}}{\sin\Theta_a}\left(
\delta_{0,j}
-2\delta_{1,j}
+\delta_{2,j}
\right)\,.
\end{equation}
The $\delta_{0,j}$ refers to an angular location ($\Phi=-\Delta\Phi/2$)
that is not on the grid. Here we can use the symmetry 
$\Psi(\Theta,-\Phi)$=$\Psi(\Theta,\Phi)$, and hence
$\Psi(\Theta,-\Delta\Phi/2)$=$\Psi(\Theta,\Delta\Phi/2)$, to replace 
$\delta_{0,j}$ with 
$\delta_{1,j}$. 
The resulting $L_{ab,ij}$ satisfies the symmetry of Eq.~(\ref{defsym}).

For $b=n_\Phi$, the considerations are very similar. The
$\Psi(\Theta,-\Phi)$=$\Psi(\Theta,\Phi)$ symmetry is used to replace
$\delta_{n_\Phi+1,j}$ by $\delta_{n_\Phi,j}$.

\subsubsection*{Case V: Boundaries at $a=n_\Theta$, $b=1$ and $b=n_\Phi$}

Here the forms of $L^{(1)}_{n_\Theta 1,ij}$ and $L^{(1)}_{n_\Theta
n_\Phi,ij}$ are taken from Case III with $b=1,n_\Phi $, and the forms
of $L^{(2)}_{n_\Theta 1,ij}$ and $L^{(2)}_{n_\Theta n_\Phi,ij}$ are
taken from Case IV with $a=n_\Theta$. From the considerations of Case
III and Case IV it follows that the results here also satisfy
Eq.~(\ref{defsym}).

To clarify the results derived above, we list here all nonzero 
elements of  $L^{(1)}_{ab,ij}$  and $L^{(2)}_{ab,ij}$: 
\begin{equation}
L^{(1)}_{ab,ab}=-\frac{\sin\Theta_{a+1/2}+\sin\Theta_{a-1/2}
}{(\Delta\Theta)^2}\quad\mbox{all $a,b$}
\end{equation}
\begin{equation}
L^{(1)}_{ab,(a-1)b}=L^{(1)}_{(a-1)b,ab}
=\frac{\sin\Theta_{a-1/2}}{(\Delta\Theta)^2}\quad
\mbox{for $1<a$}
\end{equation}
\begin{equation}
L^{(1)}_{ab,(a+1)b}=L^{(1)}_{(a+1)b,ab}
=\frac{\sin\Theta_{a+1/2}}{(\Delta\Theta)^2}\quad
\mbox{for $a<n_\Theta
$}
\end{equation}
\begin{equation}
L^{(1)}_{n_\Theta b,n_\Theta \bar{b}}=L^{(1)}_{n_\Theta \bar{b},n_\Theta {b}}
=\frac{1}{(\Delta\Theta)^2}\quad
\mbox{where  $\bar{b}=n_\Phi-b$}
\end{equation}
\begin{equation}
L^{(2)}_{a
b,ab}=-\frac{2}{\sin\Theta_a
(\Delta\Theta)^2}\quad\mbox{all $1<b<n_\Phi
$}
\end{equation}
\begin{equation}
L^{(2)}_{a1,a1}=L^{(2)}_{an_\Phi,an_\Phi
}=
-\frac{1}{\sin\Theta_a
(\Delta\Theta)^2}
\end{equation}
\begin{equation}
L^{(2)}_{ab,a(b+1)}=L^{(2)}_{a(b+1),ab}
=\frac{1}{\sin\Theta_a
(\Delta\Theta)^2}\quad\mbox{all $b<n_\Phi
$}
\end{equation}
\begin{equation}
L^{(2)}_{ab,a(b-1)}=L^{(2)}_{a(b-1),ab}
=\frac{1}{\sin\Theta_a
(\Delta\Theta)^2}\quad\mbox{all $1<b$}\,.
\end{equation}

This completes the proof that for the full range of its indices $L_{ab,ij}
$ satisfies Eq.~(\ref{defsym}). With this result in hand we can 
go on to the computation that is central
 to our eigenspectral method: finding the eigenvectors
of 
\begin{equation}\label{grideigen} 
\sum_{ij}L_{ab,ij}\,Y^{(k)}
_{ij}
=-\Lambda^{(k)}
\,\sin\Theta_a Y^{(k)}_{ab}\,,
\end{equation}
where the $k$ index indicates that the solution is the $k$th
eigensolution.  Aside from the $\sin\Theta_a$ factor on the right,
this is a standard eigenproblem for a symmetric real matrix, and we
conclude that the eigenvalues are real and the eigenvectors form a
complete basis.  It is easy to show that the factors of $\sin\Theta_a$
do not change these conclusions.

The finite difference problem in Eq.~(\ref{grideigen}), along with
Eqs.~(\ref{anglapdef}), can be seen to be the finite difference
equivalent of the continuum eigenproblem
\begin{equation}\label{conteigen} 
\nabla_{\rm ang}^2Y(\Theta,\Phi)=-\Lambda Y(\Theta,\Phi)\,.
\end{equation}
With the usual boundary conditions, the solutions of
Eq.~(\ref{conteigen}) can be taken to be the spherical harmonics, and
$\Lambda$ to have values $\ell(\ell+1)$ where $\ell $ is an
integer. The solutions of Eq.~(\ref{grideigen}) should then be
approximately proportional to the real and imaginary parts of $Y_{\ell
m}(\Theta_i,\Phi_j)$, the approximation becoming perfect as the grid
goes to the continuum limit.

We next define the inner product in the grid vector space by
the expression in Eq.~(\ref{dotdef}).
It is simple to show, following the usual pattern with eigenproblems,
that with respect to this inner product, two nondegenerate
eigenvectors $Y^{(k)}_{ij}$. and $Y^{(k')}_{ij}$ are orthogonal
as a consequence of the symmetry in 
Eq.~(\ref{defsym}).
Since
we find the grid multipoles to have no degeneracies it follows that
the solutions to Eq.~(\ref{grideigen}) constitute a complete,
orthogonal basis, and can be normalized to satisfy
Eq.~(\ref{orthonorm}). It should be clear that this is the finite
difference equivalent of well known continuum relations.  In the
continuum limit, Eq.~(\ref{dotdef}) is the inner product on the two sphere.  The
orthogonality of our grid multipoles is therefore just the finite
difference form of the orthogonality of spherical harmonics.

\end{document}